\documentclass{ieeeaccess}

\usepackage{cite}
\usepackage{amsmath,amssymb,amsfonts}
\usepackage{algorithmic}
\usepackage{graphicx}
\usepackage{textcomp}
\usepackage{booktabs}
\usepackage{multirow}

\usepackage[T1]{fontenc}
\usepackage[utf8]{inputenc}

\usepackage[table]{xcolor}
\usepackage{colortbl}
\usepackage{hhline}
\definecolor{Gray}{gray}{0.95} 
\definecolor{NavyBlue}{rgb}{0.0, 0.0, 0.5}
\definecolor{mygray}{rgb}{0.2,0.2,0.2}
\newcolumntype{g}{>{\columncolor{Gray}}c}

\newcommand{\deltacell}[1]{\cellcolor{gray!15}#1}

\usepackage{pifont}
\newcommand{\cmark}{\ding{51}} 
\newcommand{\xmark}{\ding{55}} 
\usepackage{wrapfig}
\usepackage{caption}
\usepackage{subcaption}

\usepackage{comment}
\usepackage{hyperref}
\usepackage{cleveref}

\newcommand{\etal}{\textit{et al}.}
\newcommand{\ie}{\textit{i}.\textit{e}.}
\newcommand{\eg}{\textit{e}.\textit{g}.}
\def\paper{CryoAnomaly}
\crefname{figure}{Fig.}{Figs.}

\usepackage{bm}
\makeatletter
\AtBeginDocument{\DeclareMathVersion{bold}
\SetSymbolFont{operators}{bold}{T1}{times}{b}{n}
\SetSymbolFont{NewLetters}{bold}{T1}{times}{b}{it}
\SetMathAlphabet{\mathrm}{bold}{T1}{times}{b}{n}
\SetMathAlphabet{\mathit}{bold}{T1}{times}{b}{it}
\SetMathAlphabet{\mathbf}{bold}{T1}{times}{b}{n}
\SetMathAlphabet{\mathtt}{bold}{OT1}{pcr}{b}{n}
\SetSymbolFont{symbols}{bold}{OMS}{cmsy}{b}{n}
\renewcommand\boldmath{\@nomath\boldmath\mathversion{bold}}}

\def\BibTeX{{\rm B\kern-.05em{\sc i\kern-.025em b}\kern-.08em
 T\kern-.1667em\lower.7ex\hbox{E}\kern-.125emX}}

\usepackage{atbegshi}
\AtBeginShipoutFirst{%
  \global\setbox\AtBeginShipoutBox=\hbox{\kern16pt\box\AtBeginShipoutBox}%
}

\begin{document}
\nocite{NIH,crYOLO,Topaz,CryoTransformer,CryoSegNet,Xu2025CryoMAE,Zamanos2025Cryo-EMMAE,CryoFSL,CryoGEM,CryoCCD,AdaptingSAM,UPicker}
\nocite{TopazActiveLearning,DoGPicker,Relion,APPLEPicker,SuperCryoEMPicker,DeepCryoPicker,AutoCryoPicker,UNet,SAM,MAE,MicrographCleaner,PaDiM}
\nocite{Coreset,abTEM,TEMSimulator,Multem,InsilicoTEM,VirtualIce,CryoGAN,CryoPPP,CryoSPARC}
\history{Received 30 June 2026, accepted 17 July 2026, date of publication 23 July 2026, date of current version 6 August 2026.}
\vol{14}
\year{2026}
\doi{10.1109/ACCESS.2026.3716539}

\title{CryoAnomaly: Few-Shot Cryo-EM Particle Picking via Anomaly-Guided Hard Negative Suppression}

\author{\uppercase{Riku Itsuji}\authorrefmark{1, 2}, \IEEEmembership{Student Member, IEEE},
\uppercase{Rintaro Otsubo}\authorrefmark{1, 2}, \IEEEmembership{Student Member, IEEE},
\uppercase{Ryo Fujii}\authorrefmark{1, 2}, \IEEEmembership{Member, IEEE},
\uppercase{Xingjian Li}\authorrefmark{3},
\uppercase{Xiaolong Wu}\authorrefmark{3},
\uppercase{Hideo Saito}\authorrefmark{1, 2}, \IEEEmembership{Senior Member, IEEE},
and \uppercase{Min Xu}\authorrefmark{3}}

\address[1]{Keio University, Yokohama 223-8522, Japan}
\address[2]{Keio AI Research Center, Yokohama 223-8522, Japan}
\address[3]{Carnegie Mellon University, Pittsburgh, PA 15213, USA}

\tfootnote{This work was supported in part by U.S. NSF under Grant DBI-2238093, Grant DBI-2422619, Grant IIS-2211597, and Grant MCB-2205148.}

\markboth
{Itsuji \headeretal: CryoAnomaly: Few-Shot Cryo-EM Particle Picking via Anomaly-Guided Hard Negative Suppression}
{Itsuji \headeretal: CryoAnomaly: Few-Shot Cryo-EM Particle Picking via Anomaly-Guided Hard Negative Suppression}

\corresp{Corresponding authors: Hideo Saito (hs@keio.jp) and Min Xu (mxu1@cs.cmu.edu)}

\begin{abstract}
Cryo-electron microscopy (cryo-EM) is crucial for analyzing 3D biological structures, in which automated particle picking is essential for the workflow. However, fully supervised methods require extensive manual annotations. While few-shot learning offers a potential solution, existing approaches struggle to handle the diverse contaminations inherent in real micrographs owing to insufficient negative supervision, resulting in false positives that degrade the quality of the 3D reconstruction. Although synthetic data provides abundant and perfect labels, their use has primarily been restricted to validating identical proteins or augmenting full-shot training, leaving the potential for few-shot adaptation to novel proteins unexplored. In this study, we investigate the effective utilization of synthetic data for few-shot particle picking. We identify that direct transfer fails due to a ``clean-vs-contaminated'' Sim2Real gap. To overcome this, we propose CryoAnomaly, which is a framework that turns this gap into an advantage. By employing an anomaly detector trained on clean synthetic data, we identify real-world contaminants as anomalies and suppress them via a novel anomaly-guided hard negative suppression loss. On the CryoPPP benchmark, CryoAnomaly achieves the best picking accuracy and reconstruction resolution among state-of-the-art methods in the few-shot setting. The proposed anomaly-guided loss is confirmed to be effective on datasets with diverse contamination, where reliable pseudo-anomaly masks can be generated. Our code, dataset, and project page are available at: \url{https://github.com/riku359/CryoAnomaly}, \url{https://huggingface.co/datasets/rikrikrik/CryoAnomaly}, \url{https://riku359.github.io/CryoAnomaly-page/}.
\end{abstract}


\titlepgskip=-21pt

\maketitle

\begin{figure*}[t]
 \centering
 \includegraphics[width=\linewidth]{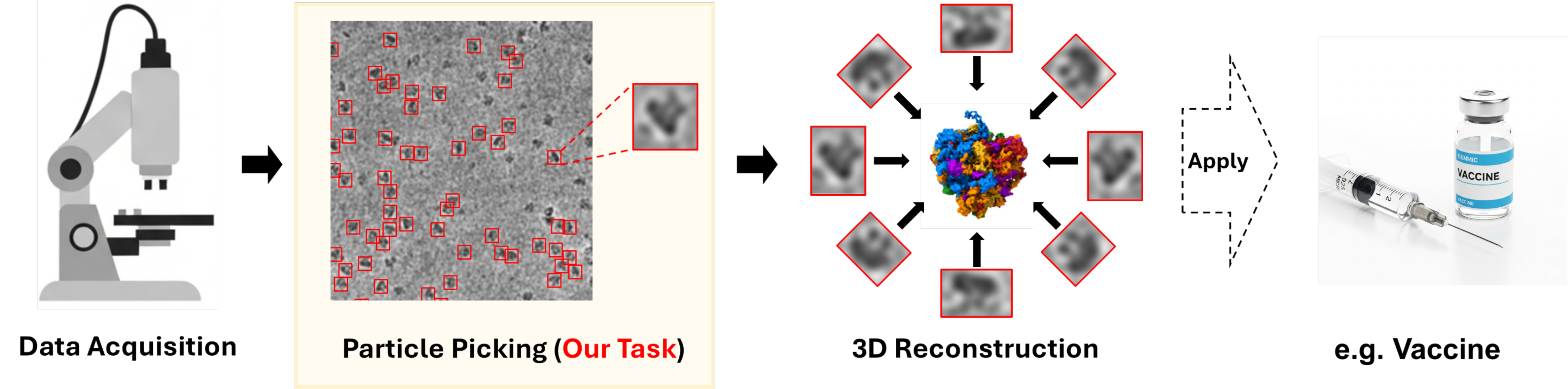}
 \caption{Overview of the cryo-EM workflow and our task. The molecular model is courtesy of NIH 3D, Entry 3DPX-003042:
\emph{Structure of the Human 80S Ribosome}~\cite{NIH}}
 \label{fig:teaser}
\end{figure*}

\section{Introduction}
\label{sec:introduction}
\PARstart{C}{}ryo-electron microscopy (cryo-EM) is a powerful technique for determining the 3D structures of cells, viruses, and protein assemblies at near-atomic resolution~\cite{Bai2015, Li2013}. This structural information is pivotal for understanding fundamental biological processes, investigating disease mechanisms~\cite{Renaud2018}, and developing novel treatments~\cite{DhakalPLI2022, Stark2022EquiBind}. As illustrated in Fig.~\ref{fig:teaser}, the cryo-EM pipeline encompasses multiple stages, among which particle picking is a pivotal step for isolating thousands of individual particle projections from 2D micrographs. Crucially, the quality of this stage significantly influences the accuracy and resolution of the reconstructed particle structures in the subsequent steps. However, achieving accurate selection is inherently difficult due to the low signal-to-noise ratio (SNR) and varied particle orientations in cryo-EM micrographs, which necessitates a large sample size for high-resolution 3D reconstructions~\cite{Frank2002, Vilas2022}. To acquire sufficient data, relying on manual picking is highly problematic.

To address this challenge, deep learning-based approaches have become the standard for automated particle picking~\cite{DeepPicker, DeepEM, FastRCNNPicker, Warp, HydraPicker, PIXER, DeepCryoPicker, McSweeney2020, DRPnet, TransPicker, CASSPER, Urdnet, LiSynergy2022, AdaptingSAM}. Fully supervised methods such as crYOLO~\cite{crYOLO}, Topaz~\cite{Topaz}, CryoTransformer~\cite{CryoTransformer}, and CryoSegNet~\cite{CryoSegNet} have achieved remarkable performance. In practice, however, constructing annotated datasets that sufficiently cover the complex and diverse conditions of macromolecular structures remains highly challenging due to the immense human cost of manual data collection and labeling. To mitigate this annotation burden, few-shot learning approaches such as CryoMAE~\cite{Xu2025CryoMAE}, Cryo-EMMAE~\cite{Zamanos2025Cryo-EMMAE}, and CryoFSL~\cite{CryoFSL} have been actively proposed. Although these methods successfully reduce data requirements, they typically underperform compared with their fully supervised counterparts. A primary bottleneck in the few-shot setting is a severe distribution gap encountered during deployment, caused by real-world contamination such as ice crystals and carbon edges. Because a limited labeled support set cannot adequately represent the diverse appearance and co-occurrence of these environmental artifacts, the models often fail to distinguish legitimate particles from high-contrast impurities. This leads to frequent false positives, which introduce invalid signals into downstream processing and negatively affect the final 3D reconstruction quality.

As a promising avenue to mitigate such distribution gaps and data scarcity, learning methods that utilize synthetic data have been widely explored in the broader computer vision community~\cite{Varol2017SURREAL,Zhang2018mixup,dunlap2023alia, kataokaijcvfractaldb,mu2020syntheticanimals, dosovitskiy2015flownet,teed2020eccv, fujii2024realtraj,Nagano2026Learning}. By leveraging synthetic data, it becomes possible to collect the variations required for generalizability at a lower cost. In cryo-EM, simulators such as CryoGEM~\cite{CryoGEM} and CryoCCD~\cite{CryoCCD} address data scarcity by generating virtually limitless micrographs with pixel-perfect ground truth. However, previous research has largely limited the use of synthetic data to validating the simulators themselves through closed-loop training and testing on identical protein structures~\cite{CryoCCD,CryoGEM}, or merely augmenting training data for fully supervised models~\cite{crYOLO, Topaz}. The use of synthetic data for few-shot adaptation to unseen proteins remains underexplored. Ideally, structural priors from simulation could be valuable in this scenario, but a distinct domain gap is a primary hurdle. Synthetic images are typically clean and lack the diverse contaminations found in the experimental data. Consequently, models pre-trained on synthetic data fail to reject real-world artifacts.

In this study, we investigate the underexplored use of synthetic data for few-shot particle picking and propose a framework named CryoAnomaly. We leverage abundant labeled synthetic data and unlabeled target-domain micrographs for the few-shot fine-tuning. Our key insight is to turn the cleanliness of synthetic data from a limitation to a strategic advantage for handling the Sim2Real gap. Instead of attempting to simulate perfect contamination, we use clean synthetic data to train an anomaly detector. When applied to real data, this detector highlights unseen contamination as an anomaly. These anomaly maps serve as pseudo-labels for a novel anomaly-guided hard negative suppression loss, which explicitly penalizes false positives on artifacts without requiring the manual annotation of negatives.
 In addition, real cryo-EM micrographs exhibit substantial image-level diversity in contrast, ice contamination, and carbon edges. Therefore, even under the same annotation budget, few-shot performance can vary depending on which micrographs are annotated. To reduce this sampling dependence, we employ a diversity-based active learning strategy that selects representative target-domain images, enabling the limited labeled support set to cover the real micrograph distribution more effectively.

Our contributions are as follows:
\begin{enumerate}
\item We investigate the underexplored use of synthetic data for few-shot particle picking, focusing on the challenge of adapting synthetic priors to unseen real-world proteins.
\item We reframe the Sim2Real gap as an anomaly detection problem and use synthetic priors to generate pseudo-anomaly masks, which guide a hard negative suppression loss to reduce false positives when reliable masks are available.
\item CryoAnomaly, our proposed few-shot pipeline that integrates synthetic supervised pre-training, diversity-based active learning, and the anomaly-guided loss, achieves the best overall performance across five CryoPPP datasets.
\end{enumerate}

\section{Related Work}
We discuss the relevant work on cryo-EM particle picking including classical methods, full-shot methods, few-shot methods, and cryo-EM simulators. To provide a clear and concise comparison, we have summarized the differences in Tab~\ref{tab:comparative overview of data usage}.

\begin{table*}[!t]
\centering
\caption{Comparative overview of data usage.}
\label{tab:comparative overview of data usage}
\renewcommand{\arraystretch}{1.2}
\resizebox{0.8\linewidth}{!}{%
 \begin{tabular}{l|c|c|c}
 \noalign{\hrule height 1pt}
 Methods & Few-shot & Unlabeled target data & Synthetic data \\ \hline
 crYOLO, Topaz~\cite{crYOLO, Topaz} & & & \cmark \\
 CryoMAE, Cryo-EMMAE, CryoFSL, He \etal~\cite{Xu2025CryoMAE, Zamanos2025Cryo-EMMAE, CryoFSL, AdaptingSAM} & \cmark & & \\
 UPicker, Kiewisz \etal~\cite{UPicker, TopazActiveLearning} & \cmark & \cmark & \\
 \rowcolor{Gray}Ours & \cmark & \cmark & \cmark \\
 \noalign{\hrule height 1pt}
 \end{tabular}%
}
\end{table*}

\subsection{Particle Picking}
Particle picking has evolved from traditional image processing techniques to deep learning methods. Early template-free methods, such as DoG Picker~\cite{DoGPicker}, can efficiently detect particles from differences in pixel intensities but risk amplifying noise. Template-based methods~\cite{Relion} improved selectivity via cross-correlation, yet they heavily rely on template quality. Classical machine learning methods employing SVMs~\cite{APPLEPicker} and clustering~\cite{SuperCryoEMPicker,DeepCryoPicker,AutoCryoPicker} have also been explored. However, their reliance on hand-crafted features proves insufficient to distinguish particle signals from complex backgrounds.

Deep learning has established a new paradigm. Unlike traditional approaches, these methods learn to recognize diverse particle characteristics and complex noise patterns directly from the training data, thereby eliminating the need for manual template design. CNN-based methods, such as crYOLO~\cite{crYOLO} and Topaz~\cite{Topaz}, are widely adopted for particle picking. crYOLO adapts the YOLO~\cite{YOLO} architecture for efficiency, whereas Topaz employs positive-unlabeled learning to handle unannotated particles in the negative regions. Transformer-based methods, such as CryoTransformer~\cite{CryoTransformer}, improve performance but incur high training costs, hindering rapid adaptation. CryoSegNet~\cite{CryoSegNet} has achieved state-of-the-art results by combining the U-Net segmentation model~\cite{UNet} and SAM~\cite{SAM}. However, such fully supervised models rely heavily on large-scale datasets, which limits their application to novel proteins.

Few-shot strategies reduce the annotation costs. Self-supervised methods such as CryoMAE~\cite{Xu2025CryoMAE} and Cryo-EMMAE~\cite{Zamanos2025Cryo-EMMAE} use Masked Autoencoders~\cite{MAE}, whereas CryoFSL~\cite{CryoFSL} and He \etal~\cite{AdaptingSAM} adapted SAM, a foundation model pre-trained on natural images. Although these approaches mitigate data scarcity, a critical challenge remains: robustness against real-world contamination. Unlike supervised models that learn to reject artifacts through extensive negative supervision, few-shot models trained on sparse positive samples often fail to distinguish particles from high-contrast impurities (\eg, ice crystals). Furthermore, these methods typically rely on random sampling for support set selection, which often fails to capture the full diversity of the data distribution.

To address contamination, UPicker~\cite{UPicker} used Micrograph Cleaner~\cite{MicrographCleaner}, but its reliance on supervised models trained on predefined classes limits its generalizability. In this study, we employ anomaly detection, treating clean synthetic data as the ``normal'' class to identify unforeseen artifacts. We utilize PaDiM~\cite{PaDiM} for its pixel-level localization, generating pseudo-masks to suppress false positives without labeled contamination data.

To overcome the limitations of random sampling, we adopt a diversity-based active learning strategy. While Kiewisz \etal~\cite{TopazActiveLearning} proposed an uncertainty-based approach, covering data diversity is more critical than refining decision boundaries in low-data regimes. We compare k-means clustering against the core-set approach~\cite{Coreset}, which selects samples farthest from the current labeled set to ensure maximum feature coverage.

\subsection{Cryo-EM Simulator}
Conventional physics-based simulators~\cite{abTEM, TEMSimulator, Multem, InsilicoTEM, VirtualIce} adhere to physical priors but often rely on simplified noise models (\eg, Gaussian), resulting in a domain gap.
To bridge this, recent deep generative models such as CryoGAN~\cite{CryoGAN}, CryoGEM~\cite{CryoGEM}, and CryoCCD~\cite{CryoCCD} synthesize realistic noise directly from real data using GANs~\cite{GAN} or Diffusion Models~\cite{Diffusion}. However, their application remains restricted to validating identical structures~\cite{CryoCCD,CryoGEM} or pre-training fully supervised models~\cite{crYOLO,Topaz}, failing to address the generalization to unseen proteins in few-shot settings (see Tab~\ref{tab:comparative overview of data usage}).

\begin{figure*}[!t]
 \centering
 \includegraphics[width=\linewidth]{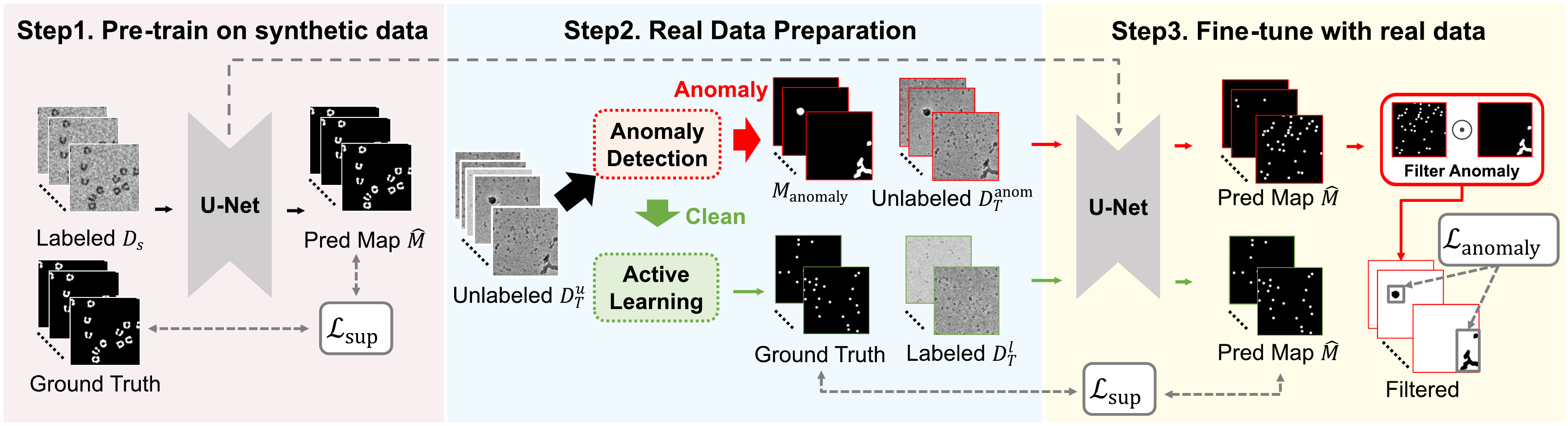}
 \caption{Overview of the CryoAnomaly pipeline. It bridges the Sim2Real gap through synthetic pre-training (Step 1), data preparation via anomaly detection and active learning (Step 2), and fine-tuning with anomaly-guided suppression (Step 3).}
 \label{fig:main_figure}
\end{figure*}

\section{Proposed Method}
\subsection{Problem Setup}
We leverage a synthetic source domain $\mathcal{D}_S$ with pixel-level annotations to facilitate learning on a target real domain. The target domain consists of a large unlabeled set $\mathcal{D}_T^u$ and a small labeled support set $\mathcal{D}_T^l$. We further define $\mathcal{D}_T^{\text{anom}} \subset \mathcal{D}_T^u$ as the subset of unlabeled data containing out-of-distribution artifacts (\eg, ice crystals), which we aim to identify and utilize for negative supervision. Unlike the source domain, $\mathcal{D}_T^l$ contains circular mask annotations derived from particle coordinates and diameter.
Our goal is to learn a mapping function $f_\theta$ to predict particle coordinates $\hat{C}$ for an unseen query set $\mathcal{Q}_T$, utilizing $\mathcal{D}_S$, $\mathcal{D}_T^{\text{anom}}$, and $\mathcal{D}_T^l$. The model predicts a probability map $\hat{M}$, which is subsequently processed by a blob detection algorithm $g(\cdot)$ to extract coordinates (\ie, $\hat{C} = g(f_\theta(x))$). Fig~\ref{fig:main_figure} illustrates our pipeline.

\subsection{Step 1: Pre-train on synthetic data}
We perform fully supervised pre-training on the synthetic dataset $\mathcal{D}_S$ using pixel-level annotations. Following CryoSegNet~\cite{CryoSegNet}, we optimize the supervised segmentation loss $\mathcal{L}_{\text{sup}}$ defined as the average of Binary Cross-Entropy and Dice loss:
\begin{equation}
 \mathcal{L}_\text{sup} = \frac{\mathcal{L}_\text{BCE} + \mathcal{L}_\text{Dice}}{2}.
\end{equation}

\subsection{Step 2: Real Data Preparation}
\begin{figure}[!ht]
 \centering
 \includegraphics[width=\linewidth]{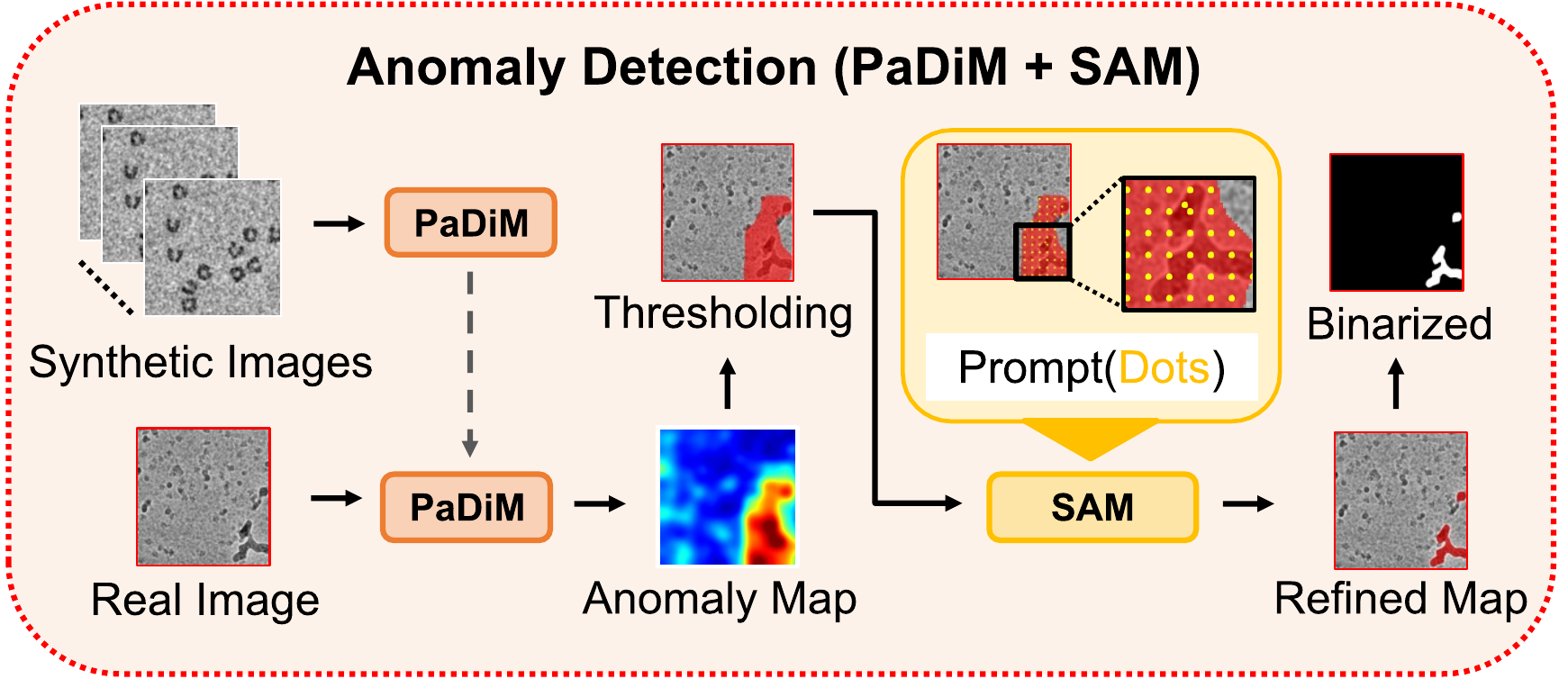}
 \caption{Overview of the anomaly mask generation pipeline. PaDiM identifies deviations from clean synthetic data, which are then refined by SAM into spatially coherent pseudo-labels for localized negative supervision.}
 \label{fig:anomaly_detection}
\end{figure}

\noindent \textbf{Anomaly Mask Creation.}
 To address real-world contaminations, we generate artifact pseudo-labels from unlabeled real micrographs $\mathcal{D}_T^u$ by integrating PaDiM~\cite{PaDiM} and SAM~\cite{SAM} (Fig~\ref{fig:anomaly_detection}). The key design is to use clean synthetic data as the normal reference and to use deviations from this reference only as negative supervision, rather than as particle labels. Specifically, we train PaDiM on clean synthetic data $\mathcal{D}_S$, defining it as the ``normal'' class. When applied to real micrographs, PaDiM produces a pixel-wise anomaly score map that coarsely localizes regions outside the synthetic distribution, such as ice contamination or carbon edges. Because these responses can be spatially coarse, we threshold the anomaly scores and refine the resulting candidate regions with SAM using grid point prompts. SAM provides class-agnostic boundary refinement, converting coarse anomaly responses into more spatially coherent masks. Finally, we apply size constraints to remove implausibly small noise regions and overly large masks, and we also reject masks larger than the corresponding PaDiM anomaly blob. The refined masks provide localized negative supervision during fine-tuning.

\begin{figure}[!ht]
 \centering
 \includegraphics[width=\linewidth]{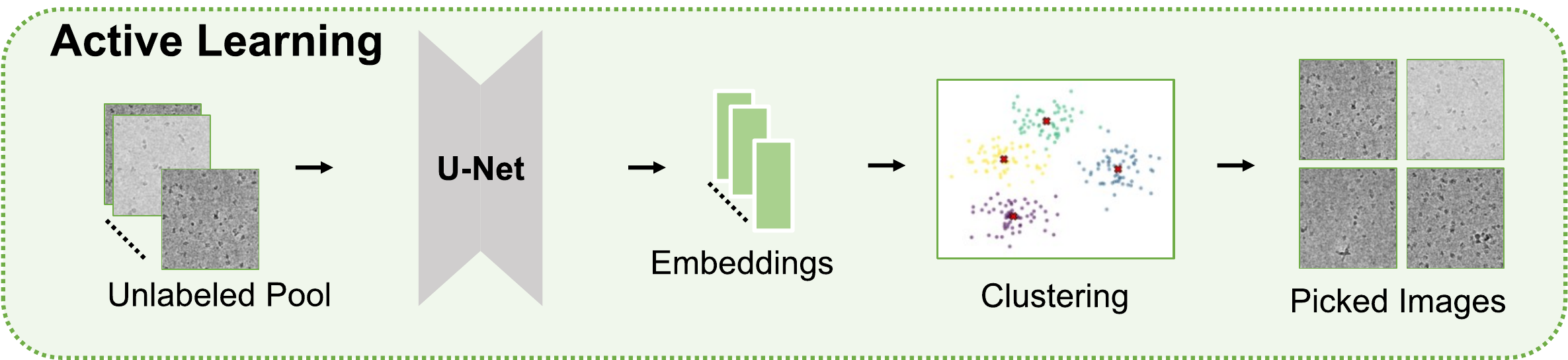}
 \caption{Diversity-based active learning strategy. By clustering U-Net embeddings, we select representative micrographs to effectively cover the target domain diversity with minimal manual annotation effort.}
 \label{fig:active_learning}
\end{figure}

\noindent \textbf{Active Learning to Select Train Images.}
To generalize to the diversity of real micrographs, we employ a diversity-based active learning strategy. We first filter out candidate micrographs with excessive anomaly areas to prioritize high-quality samples (see Fig~\ref{fig:main_figure}). For the remaining micrographs, we conduct diversity-based active learning (see Fig~\ref{fig:active_learning}). We extract embeddings using the final decoder layer of our U-Net model pre-trained on synthetic data $\mathcal{D}_S$, then select $K$ representative samples using either k-means clustering or Core-set selection~\cite{Coreset}. These strategies ensure that the limited labeled support set $\mathcal{D}_T^l$ effectively covers the target data distribution.

\subsection{Step 3: Fine-tune with real data}
We fine-tune the pre-trained model using a combination of the labeled support set $\mathcal{D}_T^l$ (selected via active learning) and the unlabeled set $\mathcal{D}_T^{\text{anom}}$ (selected via anomaly detection) with contamination pseudo-labels. For the labeled data, we employ the same segmentation loss $\mathcal{L}_{\text{sup}}$ as in the pre-training phase, ensuring the model retains its ability to detect particles.

For the unlabeled data, we introduce an anomaly-guided hard negative suppression loss $\mathcal{L}_{\text{anomaly}}$ to explicitly penalize false positives on artifacts. Using the contamination masks $M_{\text{anomaly}}$ generated by our anomaly detection module as pseudo-ground truth for the background class, we calculate BCE loss strictly within the identified contamination regions:
\begin{equation}
\mathcal{L}_\text{anomaly} = \text{BCE}(\hat{M} \odot M_\text{anomaly}, \mathbf{0}),
\end{equation}
where $\odot$ denotes pixel-wise multiplication. This forces the model to classify these areas as non-particles. Because $\mathcal{L}_\text{anomaly}$ treats all pixels inside $M_\text{anomaly}$ as background, mask refinement errors directly affect downstream learning. If refined masks mistakenly include true particles, the loss penalizes valid particle regions and can increase false negatives. Conversely, if the masks are too sparse, contaminant regions remain unsupervised and false positives may persist. Thus, the anomaly-guided loss is most effective when pseudo-anomaly masks cover contamination while avoiding valid particles. We explicitly analyze these two failure modes using the generated mask statistics and qualitative examples in Tab~\ref{tab:anomaly_mask_stats} and Fig~\ref{fig:anomaly_masks}.
The total fine-tuning objective is:
\begin{equation}
\mathcal{L}_\text{total} = \mathcal{L}_\text{sup} + \lambda \mathcal{L}_\text{anomaly},
\end{equation}
where $\lambda$ balances the suppression strength.

\subsection{Inference with test data}
\begin{figure}[!ht]
 \centering
 \includegraphics[width=\linewidth]{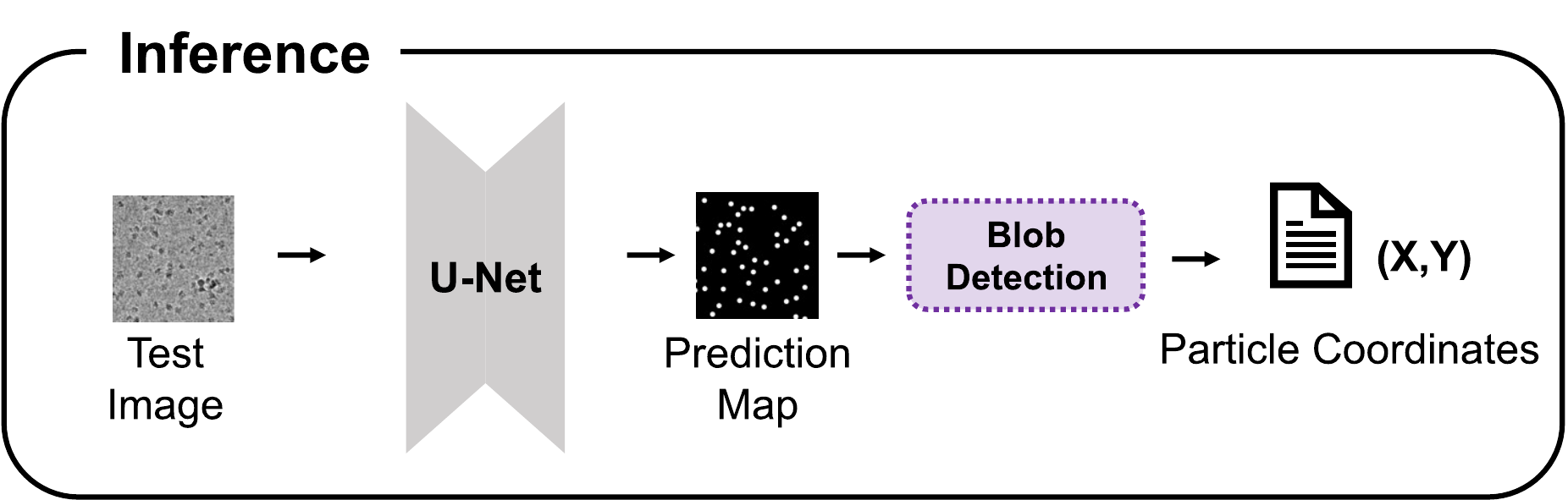}
 \caption{Inference workflow for particle picking. The model generates segmentation maps from test images, followed by blob detection to extract final coordinates for 3D reconstruction.}
 \label{fig:inference}
\end{figure}

Fig~\ref{fig:inference} illustrates the inference procedure. First, the fine-tuned model predicts segmentation probability maps for the unseen test micrographs. From these predicted maps, we extract precise particle coordinates by applying the same blob detection algorithm employed in Cryo-EMMAE~\cite{Zamanos2025Cryo-EMMAE}. These extracted coordinates are then utilized for the subsequent 3D reconstruction.

\section{Experiments}
\label{sec:experiments}

\subsection{Experimental Setup}
\label{ssec:exp_setup}

\subsubsection{Datasets}
\label{sssec:datasets}

\noindent \textbf{Source Domain (Synthetic):} We utilize synthetic micrographs from CryoCCD~\cite{CryoCCD} (EMPIAR 10005, 10061, 10199, 10289, 10425, 12667) and CryoGEM~\cite{CryoGEM} (EMPIAR 10025, 10028, 10075, 10590). We prepared 100 micrographs with pixel-level masks for each of the 10 particle types. Importantly, we ensure no overlap between source and target proteins to evaluate generalization.

\noindent \textbf{Target Domain (Real):} We evaluate five CryoPPP~\cite{CryoPPP} datasets (EMPIAR 10081, 10093, 10345, 10532, and 11056), each containing approximately 300 annotated experimental micrographs, in Tab~\ref{tab:main_results}. EMPIAR 10081 and 10345 exhibit the most diverse contamination profiles in the benchmark and thus provide the most favorable conditions for anomaly-guided suppression.

\begin{table}[t]
\centering
\caption{ Micrograph-level train/test split for each CryoPPP target dataset.}
\label{tab:data_split_protocol}
\resizebox{\linewidth}{!}{%
\begin{tabular}{c c c c}
\toprule
EMPIAR ID & Total micrographs & Train micrographs & Test micrographs \\
\midrule
10081 & 300 & 240 & 60 \\
10093 & 295 & 236 & 59 \\
10345 & 295 & 236 & 59 \\
10532 & 300 & 240 & 60 \\
11056 & 305 & 244 & 61 \\
\bottomrule
\end{tabular}%
}
\end{table}

\noindent \textbf{Full Micrograph Sets:} To assess practical reconstruction quality on all available data, we additionally use the complete micrograph sets of the two datasets with the most diverse contamination, EMPIAR 10081 (997 micrographs) and 10345 (1,644 micrographs), downloaded from the EMPIAR FTP server (\url{https://ftp.ebi.ac.uk/empiar/world_availability/}). These data do not contain the particle coordinates provided in the CryoPPP benchmark. Therefore, we cannot compute picking metrics but can evaluate 3D reconstruction quality.

\subsubsection{Baselines}
\label{sssec:baselines}
We compare CryoAnomaly against state-of-the-art methods under the same annotation budget. Baseline methods include fully supervised approaches (crYOLO~\cite{crYOLO}, Topaz~\cite{Topaz}, CryoSegNet~\cite{CryoSegNet}) and few-shot approaches (CryoFSL~\cite{CryoFSL}). CryoTransformer~\cite{CryoTransformer} was excluded because transformer-based models cannot be trained with a few labeled samples. Implementation details are provided in the Supplementary Materials.

\subsubsection{Evaluation Metrics}
\label{sssec:metrics}
We report Precision, Recall, and F1 score for particle picking results. For each ground truth particle, we identify the nearest predicted particle based on Euclidean distance. A True Positive is defined as a match where the Intersection over Union (IoU) exceeds a threshold of 0.6. Unless otherwise noted, random-support experiments share the identical support set across all compared baselines and are repeated over 10 seeds (CryoFSL: 5, due to computational cost), reported as mean$\pm$standard deviation. \paper{}'s deterministic k-means active-learning selection is evaluated once per dataset.
 Using CryoSPARC~\cite{CryoSPARC}, we also evaluate the resolution of 3D reconstructions generated from the picked particles. Because CryoSPARC reconstruction can retain stochastic variation even when the random seed is fixed, we reconstruct each identical particle stack three times and report the mean GSFSC resolution (\AA) to reduce reconstruction variance. Entries for which reconstruction was not available are marked ``-''. For the full-set reconstructions (Sec.~\ref{ssec:quant_results}), we additionally report median local resolution (\AA), estimated with CryoSPARC Local Resolution Estimation (BlocRes algorithm; FSC threshold 0.5; Adaptive Window Factor 6; remaining parameters at default), which provides a spatially resolved measure of map quality complementing the single global GSFSC value. We also report the 2D rejection rate, defined as the fraction of picked particles assigned to low-quality 2D class averages and therefore excluded from the final 3D reconstruction. Since these rejected classes are judged unsuitable for subsequent reconstruction based on their class-average quality, this rate serves as an indirect indicator of particle-stack purity. Because of their much larger particle counts, the full-set reconstructions are computed once rather than averaged over three trials. The details of the reconstruction workflow can be found in the Appendix.

\subsubsection{Data Split Protocol}
\label{sssec:data_split}
 For each target EMPIAR dataset, we first split the original micrographs into fixed training and test partitions at the micrograph level using an 80\%/20\% ratio, as summarized in Tab~\ref{tab:data_split_protocol}. Tiling is applied only after this split; therefore, tiles derived from the same original micrograph never appear in both the training and test partitions. The test partition is used only for evaluation. PaDiM is trained only on source-domain synthetic micrographs and, when applied to target data, pseudo-anomaly masks are generated only for training-partition micrographs.

In the standard 1-shot setting, one training micrograph is selected as the labeled support sample. Because \paper{} operates on tiled inputs, its 1-shot support is implemented as four non-overlapping annotated tiles from the selected training micrograph; baseline methods use the same micrograph directly without tiling. More generally, in the $N$-shot setting, $N$ training micrographs are selected as the labeled support: \paper{} tiles each selected micrograph as described above, whereas the baseline methods use the selected micrographs directly without tiling.

 For the full-set reconstruction experiments, we apply the trained models obtained from the 1-shot protocol described above. For experiments in which training is repeated with 10 random support-selection seeds, reconstruction is performed using the model trained with the first seed. Because these reconstructions use the full micrograph set, they may include a small amount of annotated training data used for 1-shot adaptation.

\subsubsection{Implementation Details}
\label{sssec:implementation}
\noindent \textbf{Synthetic Data Creation.}
We generated synthetic datasets using the official pipelines of CryoGEM~\cite{CryoGEM} and CryoCCD~\cite{CryoCCD}, following their standard protocols.

\noindent \textbf{Data Preprocessing.}
 We applied the denoising pipeline used in CryoSegNet~\cite{CryoSegNet}, consisting of standard scaling, non-local means filtering, Wiener filtering, CLAHE, and guided filtering. Standard scaling normalizes micrograph intensities and reduces low-frequency intensity variation. Non-local means filtering suppresses stochastic noise while preserving repeated local image patterns. Wiener filtering further reduces frequency-dependent noise using an estimated noise-to-signal ratio. CLAHE improves local contrast in low-SNR regions, and guided filtering smooths residual noise while preserving spatial structures. Because synthetic micrographs have weaker noise than real experimental micrographs, we used milder parameters for synthetic data to avoid over-smoothing particle structures. Excessive denoising may distort particle structures and affect the normal feature distribution learned by PaDiM. The preprocessing parameters are listed in Tab~\ref{tab:denoising_params_main}.

\begin{table}[t]
\centering
\caption{ Denoising parameters for real and synthetic data.}
\label{tab:denoising_params_main}
\resizebox{\linewidth}{!}{%
\begin{tabular}{l l c c}
\toprule
Stage & Parameter & Real & Synthetic \\
\midrule
Standard Scaling & Gaussian Kernel Size & 9 & 3 \\
Non-local Means & Filter Strength ($h$) & 10 & 3 \\
Wiener Filtering & Noise-to-Signal Ratio ($K$) & 30 & 10 \\
CLAHE & Clip Limit & 2.0 & 0.5 \\
 & Grid Size & $16 \times 16$ & $8 \times 8$ \\
Guided Filtering & Radius ($r$) & 20 & 5 \\
 & Regularization ($\epsilon$) & 0.1 & 0.05 \\
\bottomrule
\end{tabular}%
}
\end{table}

\noindent \textbf{Anomaly Mask Creation.}
We trained PaDiM using the Anomalib library with default parameters on a subset of the source domain. We used a total of 100 images composed of 10 samples from each of the 10 source categories. Anomaly maps were binarized with a threshold of 0.95. For refinement, we generated candidate masks using SAM with a $32 \times 32$ grid of prompts. We selected masks overlapping with PaDiM regions and enforced size constraints of 0.1\% to 50\% of the image area. We also rejected masks larger than the corresponding PaDiM anomaly blob.

\noindent \textbf{Network Architecture.}
We used the identical U-Net architecture as CryoSegNet~\cite{CryoSegNet}.

\noindent \textbf{Training and Inference.}
We pre-trained the model on the full source dataset (90\% train, 10\% val) for 200 epochs. We applied random cropping with a scale range of 0.45 to 0.55 and resized patches to $512 \times 512$, using a batch size of 4 and the Adam optimizer with a learning rate of $1 \times 10^{-4}$.
We fine-tuned on the target training partition for 100 epochs using the 1-shot support subset defined above and the unlabeled training micrographs with generated anomaly masks. We applied rotational augmentation ($90^{\circ}, 180^{\circ}, 270^{\circ}$) and used a batch size of 4 (2 labeled, 2 unlabeled samples) with the Adam optimizer ($1 \times 10^{-4}$). We set the anomaly loss weight $\lambda = 1.0$. At inference time, we applied the fine-tuned model only to the held-out test partition and extracted final particle coordinates using the post-processing strategy described in Cryo-EMMAE~\cite{Zamanos2025Cryo-EMMAE}.

\subsection{Quantitative Results}
\label{ssec:quant_results}

\noindent \textbf{Comparison with baselines in 1-shot setting.}
 Tab~\ref{tab:main_results} compares particle picking and 3D reconstruction results in a 1-shot setting across five CryoPPP datasets. Because our focus is few-shot adaptation, annotation-efficient methods serve as the primary baseline, while methods originally developed for fully supervised training are included as reference baselines. All methods in this table are evaluated under the same target-domain annotation budget defined in Sec.~\ref{sssec:data_split}.

CryoAnomaly achieves the highest average F1 score and best average reconstruction resolution across five datasets (AVG F1: 0.587; AVG Resolution: 9.46~\AA). The anomaly-guided loss is most effective on EMPIAR 10081 and 10345, the datasets with the most diverse contamination, where the anomaly detector produces reliable pseudo-masks. On the remaining three datasets, CryoAnomaly remains competitive or best in F1, but the margin is smaller. Although Topaz achieves the highest average recall, it picks several times more particles (AVG 52,520) and yields substantially lower precision, indicating frequent false positives that degrade 3D reconstruction quality. CryoFSL achieves competitive recall but lower F1. CryoAnomaly, by contrast, maintains a compact and precise particle set (AVG 14,283), confirming that the anomaly loss effectively suppresses contaminant-induced false positives when contamination is diverse. On EMPIAR 10345, the crYOLO test-set reconstruction shows large trial-to-trial variability (55.88$\pm$39.55~\AA), suggesting that the held-out test micrographs alone are insufficient for stable reconstruction of this dataset.

\noindent \textbf{Full-set reconstruction.} The main 1-shot comparison in Tab~\ref{tab:main_results} uses five CryoPPP datasets. Test-set-only 3D reconstruction can be unstable for some datasets and therefore is not always an appropriate basis for reconstruction comparison. We therefore additionally evaluate full-set reconstruction on EMPIAR 10081 and 10345, while limiting this analysis to these two datasets, which most clearly reveal the reconstruction-quality differences among the compared methods. We performed this analysis using CryoAnomaly, CryoSegNet, and CryoFSL, with all three methods using models trained under the identical 1-shot protocol. Results are summarized in Tab~\ref{tab:fullset}.

CryoAnomaly achieves the best GSFSC@0.143 resolution on both datasets (4.35~\AA{} and 4.47~\AA{} on EMPIAR 10081 and 10345), and the same ranking holds for median local resolution (9.30~\AA{} and 8.40~\AA{}). This is consistent with the picking results in Tab~\ref{tab:main_results}, where CryoAnomaly attains the highest average F1 with a compact particle set. As supporting metrics for the full-set reconstruction workflow, Tab~\ref{tab:fullset} reports the number of picked particles, the number retained after 2D classification, the 2D rejection rate, and the anomaly-mask overlap. The 2D rejection rate is the fraction of picked particles discarded by 2D classification as low-quality class averages before 3D reconstruction, so a lower value indicates a cleaner picked stack. CryoAnomaly's 2D rejection rate is far lower than CryoFSL's (2.7\% vs.~14.9\% on 10081; 2.4\% vs.~15.6\% on 10345) and comparable to CryoSegNet's (1.6\% and 2.1\%). As a more direct probe of stack purity, we measured the anomaly-mask overlap, the fraction of picked-particle centers that fall within the pseudo-anomaly (contamination) masks. CryoAnomaly overlaps contamination less than CryoFSL (0.07\% vs.~0.15\% on 10081; 0.02\% vs.~0.76\% on 10345); relative to CryoSegNet, the overlap is comparable on 10081 (0.07\% vs.~0.04\%) and lower on 10345 (0.02\% vs.~0.20\%). Since this overlap is computed with the same anomaly detection pipeline used during training, it is a partially circular measure and is reported only as a supporting reference; with that caveat, the lower overlap is consistent with anomaly-guided suppression reducing contaminant picks. Together with the best GSFSC and median local-resolution values, these results suggest that the CryoAnomaly pipeline, including anomaly-guided suppression, works effectively in practical full-set reconstruction settings.

\begin{table*}[t]
\centering
\caption{ 1-shot comparison on five CryoPPP datasets under the same target-domain annotation budget. Baselines are reported as mean$\pm$standard deviation. CryoAnomaly is deterministic and evaluated once. Best results in bold, second-best underlined. Resolution: 3-trial average GSFSC (\AA).}
\label{tab:main_results}
\setlength{\tabcolsep}{3pt}
\resizebox{\textwidth}{!}{%
\begin{tabular}{l l cccccc}
\toprule
Metric & Method & 10081 & 10093 & 10345 & 10532 & 11056 & AVG \\
\midrule
\multirow{5}{*}{Precision ($\uparrow$)}
 & crYOLO     & 0.592$\pm$0.185 & 0.356$\pm$0.231 & \underline{0.523$\pm$0.285} & 0.262$\pm$0.153 & 0.183$\pm$0.161 & 0.383 \\
 & Topaz      & 0.180$\pm$0.010 & 0.227$\pm$0.008 & 0.056$\pm$0.003 & 0.384$\pm$0.028 & 0.353$\pm$0.016 & 0.240 \\
 & CryoSegNet & \underline{0.601$\pm$0.086} & \underline{0.365$\pm$0.054} & 0.423$\pm$0.179 & \underline{0.522$\pm$0.042} & \underline{0.576$\pm$0.030} & \underline{0.497} \\
 & CryoFSL    & 0.329$\pm$0.027 & 0.181$\pm$0.025 & 0.138$\pm$0.024 & 0.244$\pm$0.059 & 0.276$\pm$0.012 & 0.234 \\
\rowcolor{Gray}
 & \paper{}   & \textbf{0.717} & \textbf{0.366} & \textbf{0.526} & \textbf{0.574} & \textbf{0.651} & \textbf{0.567} \\
\midrule
\multirow{5}{*}{Recall ($\uparrow$)}
 & crYOLO     & 0.594$\pm$0.228 & 0.199$\pm$0.286 & 0.291$\pm$0.242 & 0.582$\pm$0.304 & 0.733$\pm$0.034 & 0.480 \\
 & Topaz      & \textbf{0.922$\pm$0.107} & \textbf{0.967$\pm$0.030} & \underline{0.746$\pm$0.140} & \textbf{0.917$\pm$0.077} & \textbf{0.795$\pm$0.137} & \textbf{0.869} \\
 & CryoSegNet & 0.722$\pm$0.078 & 0.381$\pm$0.050 & 0.512$\pm$0.155 & 0.334$\pm$0.170 & 0.636$\pm$0.046 & 0.517 \\
 & CryoFSL    & \underline{0.853$\pm$0.083} & \underline{0.608$\pm$0.055} & \textbf{0.826$\pm$0.138} & \underline{0.664$\pm$0.090} & \underline{0.788$\pm$0.031} & \underline{0.748} \\
\rowcolor{Gray}
 & \paper{}   & 0.848 & 0.527 & 0.691 & 0.481 & 0.568 & 0.623 \\
\midrule
\multirow{5}{*}{F1 Score ($\uparrow$)}
 & crYOLO     & 0.524$\pm$0.188 & 0.116$\pm$0.076 & 0.334$\pm$0.251 & 0.276$\pm$0.099 & 0.268$\pm$0.131 & 0.304 \\
 & Topaz      & 0.300$\pm$0.011 & \underline{0.367$\pm$0.010} & 0.104$\pm$0.004 & \textbf{0.539$\pm$0.019} & 0.485$\pm$0.030 & 0.359 \\
 & CryoSegNet & \underline{0.654$\pm$0.076} & \underline{0.367$\pm$0.019} & \underline{0.415$\pm$0.112} & 0.384$\pm$0.126 & \underline{0.603$\pm$0.025} & \underline{0.485} \\
 & CryoFSL    & 0.468$\pm$0.019 & 0.275$\pm$0.021 & 0.230$\pm$0.030 & 0.346$\pm$0.060 & 0.408$\pm$0.012 & 0.345 \\
\rowcolor{Gray}
 & \paper{}   & \textbf{0.777} & \textbf{0.432} & \textbf{0.597} & \underline{0.523} & \textbf{0.607} & \textbf{0.587} \\
\midrule
\multirow{5}{*}{Resolution (\AA) ($\downarrow$)}
 & crYOLO     & 10.57$\pm$0.08 & 22.81$\pm$4.58 & 55.88$\pm$39.55 & 14.06$\pm$8.75 & - & 25.83 \\
 & Topaz      & 16.89$\pm$0.23 & 13.72$\pm$0.56 & 17.56$\pm$0.83 & 8.26$\pm$0.05 & - & 14.11 \\
 & CryoSegNet & 9.36$\pm$0.08 & \underline{8.58$\pm$0.04} & 22.31$\pm$0.47 & 5.92$\pm$0.11 & - & 11.54 \\
 & CryoFSL    & \underline{8.83$\pm$0.10} & 8.65$\pm$0.05 & \underline{17.38$\pm$0.21} & \underline{5.78$\pm$0.27} & - & \underline{10.16} \\
\rowcolor{Gray}
 & \paper{}   & \textbf{8.68} & \textbf{8.06} & \textbf{16.26} & \textbf{4.85} & - & \textbf{9.46} \\
\midrule
\multirow{5}{*}{\# Picked}
 & crYOLO     & 17,585$\pm$30,451 & 17,554$\pm$32,816 & 1,060$\pm$936 & 65,002$\pm$43,636 & 138,925$\pm$37,547 & 48,025 \\
 & Topaz      & 48,986$\pm$3,008 & 49,051$\pm$2,064 & 47,343$\pm$2,697 & 56,336$\pm$6,373 & 60,884$\pm$11,765 & 52,520 \\
 & CryoSegNet & 10,561$\pm$1,214 & 12,073$\pm$3,369 & 4,330$\pm$3,658 & 14,091$\pm$7,421 & 29,806$\pm$2,817 & 14,172 \\
 & CryoFSL    & 11,935$\pm$980 & 17,821$\pm$2,591 & 5,286$\pm$1,394 & 35,065$\pm$5,823 & 35,836$\pm$2,163 & 21,189 \\
\rowcolor{Gray}
 & \paper{}   & 10,284 & 16,069 & 3,244 & 18,343 & 23,475 & 14,283 \\
\bottomrule
\end{tabular}}
\begin{flushleft}
\footnotesize \textbf{*} 3D reconstruction parameters for EMPIAR 11056 were not available.
\par\footnotesize \textbf{*} crYOLO, Topaz, and CryoSegNet were originally proposed as fully supervised methods, whereas CryoFSL and \paper{} are few-shot methods. We evaluate them in the common 1-shot setting for a fair comparison.
\end{flushleft}
\end{table*}

\begin{table*}[t]
\centering
\caption{ Full-set reconstruction statistics on EMPIAR 10081 and 10345. All three methods use models trained under the identical 1-shot protocol. Each full-set reconstruction is computed once. Best results in bold, second-best underlined. The anomaly-mask overlap is the fraction of picked-particle centers that fall within the generated pseudo-anomaly (contamination) masks; lower values indicate a cleaner picked stack.}
\label{tab:fullset}
\resizebox{\textwidth}{!}{%
\begin{tabular}{llrrrrrrr}
\toprule
EMPIAR & Method & \# Micrographs & \# Picked & \# After 2D Classification & 2D Rejection Rate & Anomaly-Mask Overlap & GSFSC (\AA) & Median LocalRes (\AA) \\
\midrule
\multirow{3}{*}{10081}
 & CryoSegNet & 997 & 136,511 & 134,290 & 1.6\% & 0.04\% & \underline{4.52} & \underline{9.59} \\
 & CryoFSL & 997 & 210,952 & 179,577 & 14.9\% & 0.15\% & 4.53 & 9.62 \\
 \rowcolor{Gray}
 & \paper{} & 997 & 161,189 & 156,850 & 2.7\% & 0.07\% & \textbf{4.35} & \textbf{9.30} \\
\midrule
\multirow{3}{*}{10345}
 & CryoSegNet & 1,644 & 72,478 & 70,981 & 2.1\% & 0.20\% & \underline{4.90} & \underline{8.85} \\
 & CryoFSL & 1,644 & 178,751 & 150,890 & 15.6\% & 0.76\% & 7.25 & 12.23 \\
 \rowcolor{Gray}
 & \paper{} & 1,644 & 77,909 & 76,001 & 2.4\% & 0.02\% & \textbf{4.47} & \textbf{8.40} \\
\bottomrule
\end{tabular}}
\end{table*}

\begin{table*}[t]
\centering
\caption{ N-shot comparison on EMPIAR 10081 and 10345, which exhibit diverse contamination. Baseline methods are evaluated over repeated support-selection runs and reported as mean$\pm$standard deviation. CryoAnomaly is deterministic and evaluated once. The best result within each shot, dataset, and metric is bolded; the second-best is underlined.}
\label{tab:nshot_cryosegnet_cryofsl_ours_prf1_no_resolution}
\setlength{\tabcolsep}{2pt}
\resizebox{\textwidth}{!}{%
\begin{tabular}{c l ccc ccc}
\toprule
\multirow{2}{*}{Shot} & \multirow{2}{*}{Method} & \multicolumn{3}{c}{10081} & \multicolumn{3}{c}{10345} \\
\cmidrule(lr){3-5} \cmidrule(lr){6-8}
& & Prec. & Rec. & F1 & Prec. & Rec. & F1 \\
\midrule
\multirow{3}{*}{1}
 & CryoSegNet & \underline{0.601$\pm$0.086} & 0.722$\pm$0.078 & \underline{0.654$\pm$0.076} & \underline{0.423$\pm$0.179} & 0.512$\pm$0.155 & \underline{0.415$\pm$0.112} \\
 & CryoFSL    & 0.329$\pm$0.027 & \textbf{0.853$\pm$0.083} & 0.468$\pm$0.019 & 0.138$\pm$0.024 & \textbf{0.826$\pm$0.138} & 0.230$\pm$0.030 \\
\rowcolor{Gray}
 & \paper{}   & \textbf{0.717} & \underline{0.848} & \textbf{0.777} & \textbf{0.526} & \underline{0.691} & \textbf{0.597} \\
\midrule
\multirow{3}{*}{3}
 & CryoSegNet & \underline{0.705$\pm$0.023} & 0.770$\pm$0.043 & \underline{0.735$\pm$0.012} & \underline{0.462$\pm$0.115} & 0.592$\pm$0.096 & \underline{0.503$\pm$0.053} \\
 & CryoFSL    & 0.347$\pm$0.016 & \textbf{0.915$\pm$0.011} & 0.499$\pm$0.016 & 0.158$\pm$0.021 & \textbf{0.898$\pm$0.082} & 0.263$\pm$0.024 \\
\rowcolor{Gray}
 & \paper{}   & \textbf{0.773} & \underline{0.796} & \textbf{0.784} & \textbf{0.566} & \underline{0.691} & \textbf{0.622} \\
\midrule
\multirow{3}{*}{5}
 & CryoSegNet & \underline{0.746$\pm$0.029} & 0.766$\pm$0.055 & \underline{0.754$\pm$0.019} & \underline{0.523$\pm$0.095} & \underline{0.570$\pm$0.114} & \underline{0.532$\pm$0.049} \\
 & CryoFSL    & 0.355$\pm$0.014 & \textbf{0.920$\pm$0.012} & 0.509$\pm$0.013 & 0.163$\pm$0.021 & \textbf{0.905$\pm$0.051} & 0.271$\pm$0.027 \\
\rowcolor{Gray}
 & \paper{}   & \textbf{0.784} & \underline{0.776} & \textbf{0.780} & \textbf{0.610} & 0.551 & \textbf{0.579} \\
\midrule
\multirow{3}{*}{10}
 & CryoSegNet & \textbf{0.817$\pm$0.023} & \underline{0.743$\pm$0.052} & \textbf{0.776$\pm$0.020} & \textbf{0.604$\pm$0.074} & 0.582$\pm$0.103 & \underline{0.581$\pm$0.033} \\
 & CryoFSL    & 0.338$\pm$0.021 & \textbf{0.942$\pm$0.006} & 0.491$\pm$0.029 & 0.153$\pm$0.030 & \textbf{0.934$\pm$0.040} & 0.257$\pm$0.042 \\
\rowcolor{Gray}
 & \paper{}   & \underline{0.807} & \underline{0.743} & \underline{0.774} & \underline{0.591} & \underline{0.623} & \textbf{0.607} \\
\bottomrule
\end{tabular}%
}
\end{table*}

\begin{figure}[t]
\centering
\includegraphics[width=\linewidth]{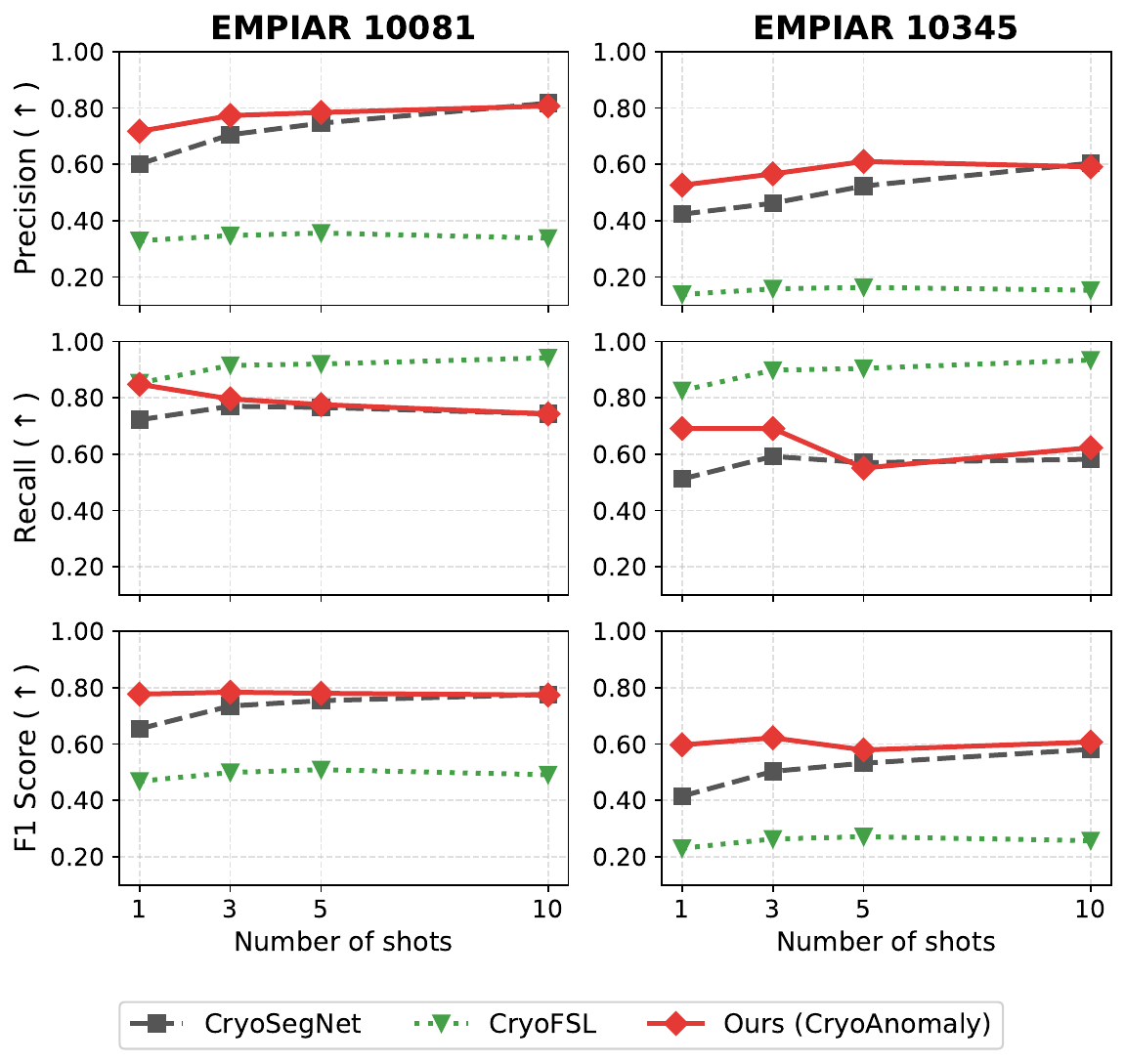}
\caption{ N-shot trajectories of Precision, Recall, and F1 Score on EMPIAR 10081 and 10345 for CryoSegNet, CryoFSL, and CryoAnomaly.}
\label{fig:nshot_prf1_3series_noresolution}
\end{figure}

\begin{table*}[!t]
\centering
\caption{Impact of synthetic pre-training on baseline models across five CryoPPP datasets. Baseline methods except CryoFSL are evaluated 10 times and reported as mean$\pm$standard deviation. CryoFSL is evaluated 5 times due to computational cost. $\Delta = \text{w/ Syn.} - \text{w/o Syn.}$; blue = improvement, red = decline. Resolution: 3-trial average GSFSC (\AA).}
\label{tab:syn_pretrain_effect}
\setlength{\tabcolsep}{2pt}
\resizebox{\textwidth}{!}{%
\begin{tabular}{l l c cccccc}
\toprule
Metric & Method & w/ Syn. & 10081 & 10093 & 10345 & 10532 & 11056 & AVG \\
\midrule
\multirow{12}{*}{Precision ($\uparrow$)}
 &     & \xmark & 0.592$\pm$0.185 & 0.356$\pm$0.231 & 0.523$\pm$0.285 & 0.262$\pm$0.153 & 0.183$\pm$0.161 & 0.383 \\
 & crYOLO & \cmark & 0.503$\pm$0.097 & 0.446$\pm$0.190 & 0.509$\pm$0.111 & 0.455$\pm$0.137 & 0.583$\pm$0.015 & 0.499 \\
 & & $\Delta$
 & \deltacell{\textcolor{red}{-0.089}}
 & \deltacell{\textcolor{blue}{+0.090}}
 & \deltacell{\textcolor{red}{-0.014}}
 & \deltacell{\textcolor{blue}{+0.193}}
 & \deltacell{\textcolor{blue}{+0.400}}
 & \deltacell{\textcolor{blue}{+0.116}} \\

 &     & \xmark & 0.180$\pm$0.010 & 0.227$\pm$0.008 & 0.056$\pm$0.003 & 0.384$\pm$0.028 & 0.353$\pm$0.016 & 0.240 \\
 & Topaz & \cmark & 0.198$\pm$0.005 & 0.255$\pm$0.010 & 0.065$\pm$0.010 & 0.400$\pm$0.019 & 0.344$\pm$0.007 & 0.251 \\
 & & $\Delta$
 & \deltacell{\textcolor{blue}{+0.018}}
 & \deltacell{\textcolor{blue}{+0.028}}
 & \deltacell{\textcolor{blue}{+0.009}}
 & \deltacell{\textcolor{blue}{+0.016}}
 & \deltacell{\textcolor{red}{-0.009}}
 & \deltacell{\textcolor{blue}{+0.011}} \\

 & & \xmark & 0.329$\pm$0.027 & 0.181$\pm$0.025 & 0.138$\pm$0.024 & 0.244$\pm$0.059 & 0.276$\pm$0.012 & 0.234 \\
 & CryoFSL & \cmark & 0.329$\pm$0.027 & 0.193$\pm$0.022 & 0.126$\pm$0.046 & 0.225$\pm$0.057 & 0.272$\pm$0.009 & 0.229 \\
 & & $\Delta$
 & \deltacell{0.000}
 & \deltacell{\textcolor{blue}{+0.012}}
 & \deltacell{\textcolor{red}{-0.012}}
 & \deltacell{\textcolor{red}{-0.019}}
 & \deltacell{\textcolor{red}{-0.004}}
 & \deltacell{\textcolor{red}{-0.005}} \\

 & & \xmark & 0.601$\pm$0.086 & 0.365$\pm$0.054 & 0.423$\pm$0.179 & 0.522$\pm$0.042 & 0.576$\pm$0.030 & 0.497 \\
 & CryoSegNet & \cmark & 0.673$\pm$0.067 & 0.379$\pm$0.031 & 0.487$\pm$0.108 & 0.530$\pm$0.043 & 0.589$\pm$0.015 & 0.532 \\
 & & $\Delta$
 & \deltacell{\textcolor{blue}{+0.072}}
 & \deltacell{\textcolor{blue}{+0.014}}
 & \deltacell{\textcolor{blue}{+0.064}}
 & \deltacell{\textcolor{blue}{+0.008}}
 & \deltacell{\textcolor{blue}{+0.013}}
 & \deltacell{\textcolor{blue}{+0.035}} \\
\midrule

\multirow{12}{*}{Recall ($\uparrow$)}
 & & \xmark & 0.594$\pm$0.228 & 0.199$\pm$0.286 & 0.291$\pm$0.242 & 0.582$\pm$0.304 & 0.733$\pm$0.034 & 0.480 \\
 & crYOLO & \cmark & 0.788$\pm$0.116 & 0.367$\pm$0.327 & 0.688$\pm$0.192 & 0.397$\pm$0.352 & 0.700$\pm$0.183 & 0.588 \\
 & & $\Delta$
 & \deltacell{\textcolor{blue}{+0.194}}
 & \deltacell{\textcolor{blue}{+0.168}}
 & \deltacell{\textcolor{blue}{+0.397}}
 & \deltacell{\textcolor{red}{-0.185}}
 & \deltacell{\textcolor{red}{-0.033}}
 & \deltacell{\textcolor{blue}{+0.108}} \\

 & & \xmark & 0.922$\pm$0.107 & 0.967$\pm$0.030 & 0.746$\pm$0.140 & 0.917$\pm$0.077 & 0.795$\pm$0.137 & 0.869 \\
 & Topaz & \cmark & 0.967$\pm$0.043 & 0.959$\pm$0.017 & 0.693$\pm$0.243 & 0.971$\pm$0.018 & 0.961$\pm$0.027 & 0.910 \\
 & & $\Delta$
 & \deltacell{\textcolor{blue}{+0.045}}
 & \deltacell{\textcolor{red}{-0.008}}
 & \deltacell{\textcolor{red}{-0.053}}
 & \deltacell{\textcolor{blue}{+0.054}}
 & \deltacell{\textcolor{blue}{+0.166}}
 & \deltacell{\textcolor{blue}{+0.041}} \\

 & & \xmark & 0.853$\pm$0.083 & 0.608$\pm$0.055 & 0.826$\pm$0.138 & 0.664$\pm$0.090 & 0.788$\pm$0.031 & 0.748 \\
 & CryoFSL & \cmark & 0.853$\pm$0.083 & 0.581$\pm$0.050 & 0.836$\pm$0.146 & 0.696$\pm$0.072 & 0.793$\pm$0.034 & 0.752 \\
 & & $\Delta$
 & \deltacell{0.000}
 & \deltacell{\textcolor{red}{-0.027}}
 & \deltacell{\textcolor{blue}{+0.010}}
 & \deltacell{\textcolor{blue}{+0.032}}
 & \deltacell{\textcolor{blue}{+0.005}}
 & \deltacell{\textcolor{blue}{+0.004}} \\

 & & \xmark & 0.722$\pm$0.078 & 0.381$\pm$0.050 & 0.512$\pm$0.155 & 0.334$\pm$0.170 & 0.636$\pm$0.046 & 0.517 \\
 & CryoSegNet & \cmark & 0.797$\pm$0.061 & 0.442$\pm$0.027 & 0.582$\pm$0.110 & 0.400$\pm$0.162 & 0.662$\pm$0.041 & 0.577 \\
 & & $\Delta$
 & \deltacell{\textcolor{blue}{+0.075}}
 & \deltacell{\textcolor{blue}{+0.061}}
 & \deltacell{\textcolor{blue}{+0.070}}
 & \deltacell{\textcolor{blue}{+0.066}}
 & \deltacell{\textcolor{blue}{+0.026}}
 & \deltacell{\textcolor{blue}{+0.060}} \\
\midrule

\multirow{12}{*}{F1 Score ($\uparrow$)}
 & & \xmark & 0.524$\pm$0.188 & 0.116$\pm$0.076 & 0.334$\pm$0.251 & 0.276$\pm$0.099 & 0.268$\pm$0.131 & 0.304 \\
 & crYOLO & \cmark & 0.596$\pm$0.086 & 0.258$\pm$0.158 & 0.543$\pm$0.082 & 0.335$\pm$0.244 & 0.623$\pm$0.094 & 0.471 \\
 & & $\Delta$
 & \deltacell{\textcolor{blue}{+0.072}}
 & \deltacell{\textcolor{blue}{+0.142}}
 & \deltacell{\textcolor{blue}{+0.209}}
 & \deltacell{\textcolor{blue}{+0.059}}
 & \deltacell{\textcolor{blue}{+0.355}}
 & \deltacell{\textcolor{blue}{+0.167}} \\

 & & \xmark & 0.300$\pm$0.011 & 0.367$\pm$0.010 & 0.104$\pm$0.004 & 0.539$\pm$0.019 & 0.485$\pm$0.030 & 0.359 \\
 & Topaz & \cmark & 0.328$\pm$0.008 & 0.402$\pm$0.013 & 0.116$\pm$0.006 & 0.567$\pm$0.017 & 0.507$\pm$0.007 & 0.383 \\
 & & $\Delta$
 & \deltacell{\textcolor{blue}{+0.028}}
 & \deltacell{\textcolor{blue}{+0.035}}
 & \deltacell{\textcolor{blue}{+0.012}}
 & \deltacell{\textcolor{blue}{+0.028}}
 & \deltacell{\textcolor{blue}{+0.022}}
 & \deltacell{\textcolor{blue}{+0.024}} \\

 & & \xmark & 0.468$\pm$0.019 & 0.275$\pm$0.021 & 0.230$\pm$0.030 & 0.346$\pm$0.060 & 0.408$\pm$0.012 & 0.345 \\
 & CryoFSL & \cmark & 0.468$\pm$0.019 & 0.287$\pm$0.018 & 0.209$\pm$0.072 & 0.333$\pm$0.065 & 0.404$\pm$0.008 & 0.340 \\
 & & $\Delta$
 & \deltacell{0.000}
 & \deltacell{\textcolor{blue}{+0.012}}
 & \deltacell{\textcolor{red}{-0.021}}
 & \deltacell{\textcolor{red}{-0.013}}
 & \deltacell{\textcolor{red}{-0.004}}
 & \deltacell{\textcolor{red}{-0.005}} \\

 & & \xmark & 0.654$\pm$0.076 & 0.367$\pm$0.019 & 0.415$\pm$0.112 & 0.384$\pm$0.126 & 0.603$\pm$0.025 & 0.485 \\
 & CryoSegNet & \cmark & 0.725$\pm$0.040 & 0.407$\pm$0.022 & 0.512$\pm$0.043 & 0.437$\pm$0.115 & 0.622$\pm$0.014 & 0.541 \\
 & & $\Delta$
 & \deltacell{\textcolor{blue}{+0.071}}
 & \deltacell{\textcolor{blue}{+0.040}}
 & \deltacell{\textcolor{blue}{+0.097}}
 & \deltacell{\textcolor{blue}{+0.053}}
 & \deltacell{\textcolor{blue}{+0.019}}
 & \deltacell{\textcolor{blue}{+0.056}} \\
\midrule

\multirow{12}{*}{Resolution (\AA) ($\downarrow$)}
 & & \xmark & 10.57$\pm$0.08 & 22.81$\pm$4.58 & 55.88$\pm$39.55 & 14.06$\pm$8.75 & - & 25.83 \\
 & crYOLO & \cmark & 10.70$\pm$0.12 & 13.40$\pm$1.16 & 18.08$\pm$0.61 & 10.64$\pm$0.10 & - & 13.21 \\
 & & $\Delta$
 & \deltacell{\textcolor{red}{+0.13}}
 & \deltacell{\textcolor{blue}{-9.41}}
 & \deltacell{\textcolor{blue}{-37.80}}
 & \deltacell{\textcolor{blue}{-3.42}}
 & \deltacell{-}
 & \deltacell{\textcolor{blue}{-12.62}} \\

 & & \xmark & 16.89$\pm$0.23 & 13.72$\pm$0.56 & 17.56$\pm$0.83 & 8.26$\pm$0.04 & - & 14.11 \\
 & Topaz & \cmark & 17.27$\pm$0.17 & 12.13$\pm$0.38 & 17.90$\pm$0.68 & 9.24$\pm$2.77 & - & 14.13 \\
 & & $\Delta$
 & \deltacell{\textcolor{red}{+0.38}}
 & \deltacell{\textcolor{blue}{-1.59}}
 & \deltacell{\textcolor{red}{+0.34}}
 & \deltacell{\textcolor{red}{+0.98}}
 & \deltacell{-}
 & \deltacell{\textcolor{red}{+0.03}} \\

 & & \xmark & 8.83$\pm$0.10 & 8.65$\pm$0.05 & 17.38$\pm$0.21 & 5.78$\pm$0.27 & - & 10.16 \\
 & CryoFSL & \cmark & 8.84$\pm$0.03 & 8.66$\pm$0.05 & 17.40$\pm$0.21 & 5.47$\pm$0.05 & - & 10.09 \\
 & & $\Delta$
 & \deltacell{\textcolor{red}{+0.01}}
 & \deltacell{\textcolor{red}{+0.01}}
 & \deltacell{\textcolor{red}{+0.02}}
 & \deltacell{\textcolor{blue}{-0.31}}
 & \deltacell{-}
 & \deltacell{\textcolor{blue}{-0.07}} \\

 & & \xmark & 9.36$\pm$1.13 & 8.58$\pm$0.15 & 22.31$\pm$10.75 & 5.92$\pm$1.48 & - & 11.54 \\
 & CryoSegNet & \cmark & 8.81$\pm$0.23 & 8.36$\pm$0.11 & 17.68$\pm$1.42 & 5.54$\pm$1.53 & - & 10.10 \\
 & & $\Delta$
 & \deltacell{\textcolor{blue}{-0.55}}
 & \deltacell{\textcolor{blue}{-0.22}}
 & \deltacell{\textcolor{blue}{-4.63}}
 & \deltacell{\textcolor{blue}{-0.38}}
 & \deltacell{-}
 & \deltacell{\textcolor{blue}{-1.44}} \\
\bottomrule
\end{tabular}}
\end{table*}

\begin{table*}[t]
\centering
\caption{ Integrated ablation of tiling and algorithmic components on EMPIAR 10081 and 10345 in the 1-shot setting. Stochastic rows are evaluated 10 times and reported as mean$\pm$standard deviation. Active-learning rows are deterministic and evaluated once. Resolution: 3-trial average GSFSC (\AA). Best results in bold, second-best underlined.}
\label{tab:ablation_study_out_method_components}
\resizebox{0.8\linewidth}{!}{%
\begin{tabular}{ccccc cc cc}
\toprule
\multirow{2}{*}{FT} &
\multirow{2}{*}{Tiling} &
\multirow{2}{*}{Pretrain} &
\multirow{2}{*}{Active Learning} &
\multirow{2}{*}{Anomaly Loss} &
\multicolumn{2}{c}{F1 Score ($\uparrow$)} &
\multicolumn{2}{c}{Resolution (\AA) ($\downarrow$)} \\
\cmidrule(lr){6-7} \cmidrule(lr){8-9}
& & & & & 10081 & 10345 & 10081 & 10345 \\
\midrule
\cmark & & & & & 0.654$\pm$0.076 & 0.415$\pm$0.112 & 9.36$\pm$1.13 & 22.31$\pm$10.75 \\
\cmark & \cmark & & & & 0.648$\pm$0.053 & 0.455$\pm$0.077 & 9.11$\pm$0.42 & 18.14$\pm$1.94 \\
\cmark & \cmark & \cmark & & & 0.755$\pm$0.026 & 0.559$\pm$0.023 & 8.83$\pm$0.23 & 18.59$\pm$4.53 \\
\cmark & \cmark & \cmark & \cmark & & \underline{0.767} & 0.555 & \underline{8.72} & \underline{16.84} \\
\cmark & \cmark & \cmark & & \cmark & 0.766$\pm$0.010 & \underline{0.566$\pm$0.015} & 8.79$\pm$0.19 & 17.93$\pm$1.36 \\
\cmark & \cmark & \cmark & \cmark & \cmark & \textbf{0.777} & \textbf{0.597} & \textbf{8.68} & \textbf{16.26} \\
\bottomrule
\end{tabular}%
}
\end{table*}

\begin{table}[t]
\centering
\caption{ Ablation study of pre-training synthetic data. Resolution: 3-trial average GSFSC (\AA). Best results in bold, second-best underlined.}
\label{tab:ablation_study_pretrain_data}
\resizebox{\linewidth}{!}{%
 \begin{tabular}{l ccc ccc}
 \toprule
 \multirow{2}{*}{Method} &
 \multicolumn{3}{c}{F1 Score ($\uparrow$)} &
 \multicolumn{3}{c}{Resolution (\AA) ($\downarrow$)} \\
 \cmidrule(lr){2-4} \cmidrule(lr){5-7}
 & 10081 & 10345 & Avg. & 10081 & 10345 & Avg. \\
 \midrule
 No pretrain & 0.648 & 0.455 & 0.551 & 9.11 & 18.14 & 13.62 \\
 CryoGEM & 0.734 & 0.523 & 0.629 & \underline{8.73} & \underline{17.05} & \underline{12.89} \\
 CryoCCD & \textbf{0.778} & \underline{0.591} & \underline{0.684} & 8.91 & 17.28 & 13.10 \\
 GEM + CCD & \underline{0.777} & \textbf{0.597} & \textbf{0.687} & \textbf{8.68} & \textbf{16.26} & \textbf{12.47} \\
 \bottomrule
 \end{tabular}%
}
\end{table}

\begin{table*}[t]
\centering
\caption{ Effectiveness of active learning. Random and CoreSet are evaluated 10 times and reported as mean$\pm$standard deviation. k-means is deterministic and evaluated once. Resolution: 3-trial average GSFSC (\AA). Best results in bold, second-best underlined.}
\label{tab:active-learning-ablation-metrics}
\resizebox{0.8\linewidth}{!}{%
 \begin{tabular}{l ccc ccc}
 \toprule
 \multirow{2}{*}{Method} &
 \multicolumn{3}{c}{F1 Score ($\uparrow$)} &
 \multicolumn{3}{c}{Resolution (\AA) ($\downarrow$)} \\
 \cmidrule(lr){2-4} \cmidrule(lr){5-7}
 & 10081 & 10345 & Avg. & 10081 & 10345 & Avg. \\
 \midrule
 Random & \underline{0.766$\pm$0.010} & \underline{0.566$\pm$0.015} & \underline{0.666} & \underline{8.79$\pm$0.11} & \underline{17.94$\pm$0.78} & \underline{13.37$\pm$0.40} \\
 CoreSet & 0.751$\pm$0.012 & 0.529$\pm$0.079 & 0.640 & 8.86$\pm$0.12 & 25.92$\pm$22.42 & 17.39$\pm$11.21 \\
 k-means & \textbf{0.777} & \textbf{0.597} & \textbf{0.687} & \textbf{8.68} & \textbf{16.26} & \textbf{12.47} \\
 \bottomrule
 \end{tabular}%
}
\end{table*}

\begin{table}[t]
\centering
\caption{ Comparison of F1 score and Resolution (\AA) across different input tiling scales. The "1/n" scale denotes dividing the micrograph into an $n \times n$ grid before resizing to the model input size. Resolution: 3-trial average GSFSC (\AA). Best results in bold, second-best underlined.}
\label{tab:resolution-ablation-metrics}
\resizebox{\linewidth}{!}{%
 \begin{tabular}{l ccc ccc}
 \toprule
 \multirow{2}{*}{Input Scale} &
 \multicolumn{3}{c}{F1 Score} &
 \multicolumn{3}{c}{Resolution (\AA)} \\
 \cmidrule(lr){2-4} \cmidrule(lr){5-7}
 & 10081 & 10345 & Avg. & 10081 & 10345 & Avg. \\
 \midrule
 $original$ & 0.640 & 0.380 & 0.510 & 9.01 & \underline{17.96} & \underline{13.49} \\
 $1/2$ & \textbf{0.777} & \textbf{0.597} & \textbf{0.687} & \textbf{8.68} & \textbf{16.26} & \textbf{12.47} \\
 $1/3$ & \underline{0.753} & \underline{0.527} & \underline{0.640} & \underline{8.70} & 18.59 & 13.65 \\
 \bottomrule
 \end{tabular}%
}
\end{table}

\begin{table}[!t]
 \caption{ Ablation study of anomaly loss weight $\lambda$ on EMPIAR 10081 and 10345. F1 Score and Resolution are reported as the average over the two datasets; Resolution: 3-trial average GSFSC (\AA).}
\begin{center}
\resizebox{0.8\linewidth}{!}{
\begin{tabular}{ccccc}
\toprule
\multirow{2}{*}{Metric} & \multicolumn{4}{c}{$\lambda$} \\
 \cmidrule(l{2pt}r{3pt}){2-5}
 & 0.5 & 0.75 & 1.0 & 1.25 \\
 \midrule
F1 Score & 0.666 & 0.679 & \textbf{0.687} & 0.678 \\
Resolution (\AA) & 13.31 & 12.75 & 12.47 & \textbf{12.04} \\
\bottomrule
\end{tabular}
}
\vspace{-1.em}
\end{center}
 \label{tab:ablation_anomaly_loss_weight}
\end{table}

\noindent \textbf{N-shot Analysis.}
The ablation studies and the N-shot analysis are conducted on EMPIAR 10081 and 10345, the datasets with diverse contamination on which our pipeline is designed to be effective. We evaluate performance across annotation budgets of 1, 3, 5, and 10 shots. Tab~\ref{tab:nshot_cryosegnet_cryofsl_ours_prf1_no_resolution} reports Precision, Recall, and F1 Score for CryoSegNet, CryoFSL, and CryoAnomaly across all annotation budgets. Fig~\ref{fig:nshot_prf1_3series_noresolution} visualizes the corresponding trajectories. CryoAnomaly outperforms the baseline methods particularly in the low-shot regime, and its advantage is larger when the number of shots is smaller.

\subsection{Generalizability of Synthetic Priors across Architectures}
\label{ssec:generalizability_syn_priors}
To evaluate the generalizability of synthetic data as a structural prior, we examine the impact of synthetic pre-training on various baseline architectures independently of our proposed components. Tab~\ref{tab:syn_pretrain_effect} indicates that architectures such as crYOLO, Topaz, and CryoSegNet benefit from synthetic initialization, exhibiting an overall improvement in picking sensitivity and F1 scores. Conversely, CryoFSL shows minimal sensitivity to synthetic data, as its reliance on a frozen SAM~\cite{SAM} encoder limits the network's capacity to internalize domain-specific features from the simulation. However, despite the gains in recall, we observed a degradation in 3D resolution for Topaz and a decrease in precision for crYOLO. These results suggest that while synthetic pre-training provides robust structural knowledge, it remains insufficient for distinguishing true particles from high-contrast real-world contaminants. Consequently, synthetic pre-training serves as a powerful yet incomplete solution, underscoring the necessity of the specialized suppression mechanisms introduced in our full framework to bridge the remaining Sim2Real gap.

\subsection{Ablation Studies}
\noindent \textbf{Components of Our Framework.}
 Tab~\ref{tab:ablation_study_out_method_components} presents an integrated ablation of tiling and the three algorithmic components on EMPIAR 10081 and 10345 in the 1-shot setting. Tiling is a preprocessing step of the CryoAnomaly pipeline; we include it as an explicit ablation factor so that its contribution can be read alongside the algorithmic components, but it is a generic preprocessing choice rather than a distinguishing methodological contribution of CryoAnomaly.

The first two rows isolate the effect of tiling. Adding tiling while keeping all other components disabled leaves the F1 score almost unchanged on EMPIAR 10081 (0.654 to 0.648), but improves it on 10345 (0.415 to 0.455). This suggests that tiling is particularly beneficial when the model needs more local image contexts from the same limited annotation budget.

Adding synthetic pre-training on top of tiling provides the largest F1 gain on both datasets, confirming the value of structural priors from pixel-accurate synthetic masks. When active learning or anomaly loss is added individually to the tiled and pre-trained model, both improve the results and yield comparable F1 scores on EMPIAR 10081 (0.767 and 0.766). Using both components together achieves the best performance across both datasets and both metrics, showing that targeted sample selection and contaminant suppression are complementary.

\noindent \textbf{Choice of Synthetic Pre-training Data.}
Tab~\ref{tab:ablation_study_pretrain_data} demonstrates that while any synthetic pre-training outperforms the non-pretrained baseline, combining datasets (GEM + CCD) yields the best overall performance (Avg F1: 0.687, Resolution: 12.47~\AA), suggesting that data diversity aids generalization. However, single-source pre-training occasionally excels on individual datasets (\eg, CryoCCD attains the highest F1 on EMPIAR 10081), indicating that distributional similarity between synthetic and real data remains a relevant factor for optimization.

\noindent \textbf{Impact of Active Learning Strategy.}
Tab~\ref{tab:active-learning-ablation-metrics} shows that k-means clustering achieves the highest average F1 (0.687) among the three selection strategies. Unlike Core-set selection, which prioritizes outliers and tends to pick uninformative edge cases given the extremely limited budget (4 patches), k-means effectively captures representative central tendencies, making it better suited for few-shot adaptation under this budget.

\subsection{Sensitivity Analysis}

\noindent \textbf{Input image scale.}
As shown in Tab~\ref{tab:resolution-ablation-metrics}, the tiling strategy consistently outperforms the naive use of the full original micrograph. The strategy effectively maintains the relative resolution of particles fed into the network, preventing the loss of high-frequency details caused by the downsampling process in the U-Net encoder.
Specifically, the $1/2$ scale achieves the optimal balance, recording the highest average F1 score of 0.687 and the best (lowest) 3D reconstruction resolution of 12.47~\AA.
While the $1/3$ scale increases the effective magnification, which can aid the detection of low-contrast particles, it underperforms the $1/2$ scale on both datasets with diverse contamination.
This degradation at $1/3$ scale can be attributed to the loss of global contextual information (\eg, difficulty in recognizing large contamination patterns) and the overlooking of particles on the tile boundaries.
Based on these findings, we identify the $1/2$ scale as the optimal setting for our framework.

\noindent \textbf{Anomaly loss weights.}
We analyze the impact of the hyperparameter $\lambda$, which balances the contribution of the anomaly loss $\mathcal{L}_{\text{anomaly}}$ relative to the supervision loss $\mathcal{L}_{\text{sup}}$. As shown in Tab~\ref{tab:ablation_anomaly_loss_weight}, $\lambda=1$ records the best average F1 score of 0.687. Using a smaller $\lambda$ (\eg, 0.5) fails to sufficiently penalize false positives in contaminated regions, leading to reduced precision. Conversely, setting $\lambda$ too high (\eg, 1.25) causes the regularization term to dominate the loss, destabilizing the learning process and degrading F1. Although the average reconstruction resolution is marginally better at $\lambda=1.25$ (12.04~\AA{} vs.~12.47~\AA{}), the resolution metric is far less stable than F1 across datasets, so we rely on the more robust F1 trend.
Therefore, we select $\lambda = 1.0$ as the optimal trade-off.

\begin{figure}[!t]
 \centering
 \includegraphics[width=\linewidth]{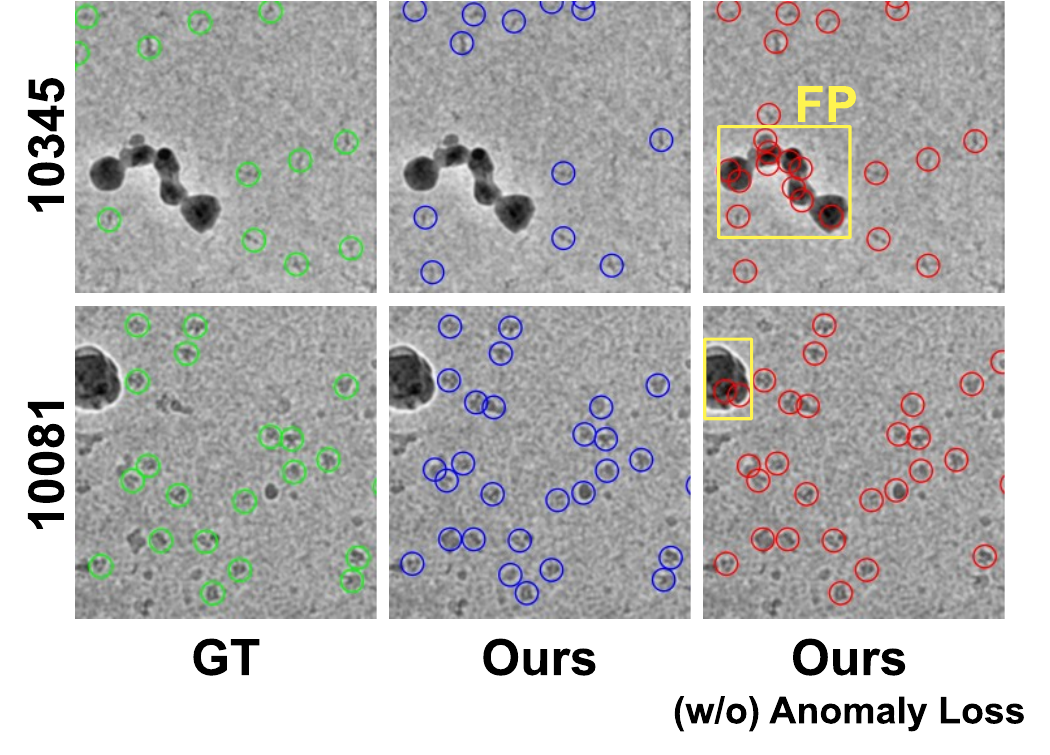}
 \caption{Qualitative comparison of particle picking with and without the anomaly-guided loss.
Green, blue, and red circles represent the ground truth (GT), predictions by our method, and predictions without (w/o) anomaly loss, respectively. The yellow boxes highlight False Positives (FP) on contaminants, which are prominent when the anomaly loss is excluded.}
 \label{fig:qualitative_picking}
\end{figure}

\subsection{Qualitative Evaluation}
\label{ssec:qualitative}

\noindent \textbf{2D picking comparison.}
Fig~\ref{fig:qualitative_picking} visualizes picking results comparing our full method against a variant without anomaly loss. The baseline model (without anomaly loss) mistakenly classifies high-contrast contaminants as particles (False Positives, yellow boxes). In contrast, our method effectively suppresses these artifacts, demonstrating that the anomaly-guided loss successfully teaches the model to distinguish true particles from domain-specific hard negatives.

\begin{figure}[t]
\centering
\includegraphics[width=\linewidth]{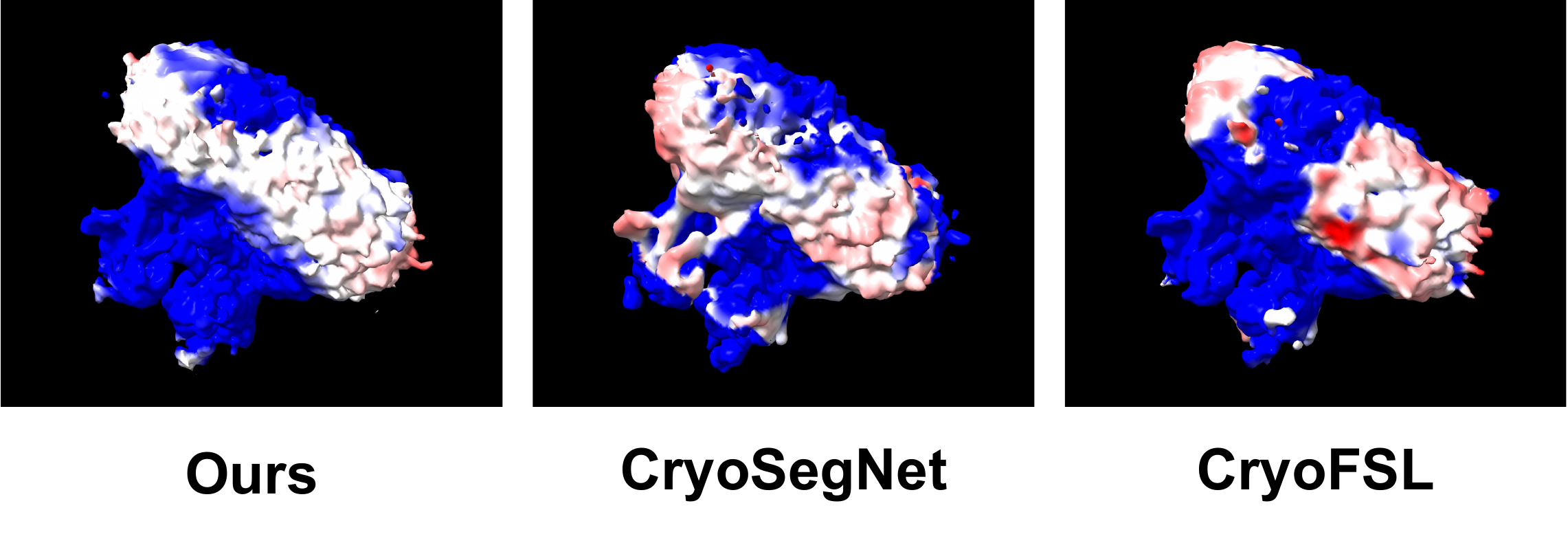}
\caption{ Local resolution maps of the full-set 3D reconstructions on EMPIAR 10081 for CryoAnomaly, CryoSegNet, and CryoFSL. Surface rendering uses the sharpened density maps; coloring is derived from the local resolution maps computed by CryoSPARC Local Resolution Estimation. Blue indicates finer local resolution, whereas red indicates coarser local resolution. The CryoFSL and CryoSegNet maps were pre-aligned to CryoAnomaly using CryoSPARC Align 3D Maps. All panels share a common color scale. Median local resolution values are annotated on each panel.}
\label{fig:local_resolution}
\end{figure}

\noindent \textbf{3D local-resolution visualization.}
Fig~\ref{fig:local_resolution} shows local resolution maps of the full-set reconstructions on EMPIAR 10081 for the three methods. In these maps, blue indicates finer local resolution (lower \AA), whereas red indicates coarser local resolution (higher \AA). CryoAnomaly shows broader blue-to-cyan regions and fewer red-to-orange coarse regions than the baselines. CryoSegNet contains more coarse patches, and CryoFSL shows the largest low-resolution regions. This spatial pattern is consistent with the global GSFSC and median local resolution values in Tab~\ref{tab:fullset}.

\begin{table*}[t]
\centering
\caption{ Analysis of generated anomaly masks and their effect on performance. Mistake Rate is the fraction of masked images whose masks erroneously overlap true particles. For F1 and Resolution we report the values without and with the anomaly loss side by side, and the better value is bolded. Resolution: 3-trial average GSFSC (\AA).}
\label{tab:anomaly_mask_stats}
\resizebox{\textwidth}{!}{%
\begin{tabular}{c c c c c c cc cc l}
\toprule
\multirow{2}{*}{Case} & \multirow{2}{*}{Dataset} & \multirow{2}{*}{Train Images} &
\multirow{2}{*}{\# Masked Images} & \multirow{2}{*}{\# Mistaken Images} &
\multirow{2}{*}{Mistake Rate} &
\multicolumn{2}{c}{F1 Score ($\uparrow$)} & \multicolumn{2}{c}{Avg. Res.~(\AA) ($\downarrow$)} &
\multirow{2}{*}{Mask regime} \\
\cmidrule(lr){7-8} \cmidrule(lr){9-10}
 & & & & & & w/o anomaly & w/ anomaly & w/o anomaly & w/ anomaly & \\
\midrule
\multirow{2}{*}{Good case}
& 10081 & 240 & 31 & 0 & 0.0\% & 0.767 & \textbf{0.777} & 8.72 & \textbf{8.68} & Reliable \\
& 10345 & 236 & 96 & 0 & 0.0\% & 0.555 & \textbf{0.597} & 16.84 & \textbf{16.26} & Reliable \\
\midrule
\multirow{3}{*}{Bad case}
& 10093 & 236 & 3 & 0 & 0.0\% & \textbf{0.465} & 0.432 & 8.26 & \textbf{8.06} & Too sparse \\
& 10532 & 240 & 188 & 38 & 20.2\% & \textbf{0.563} & 0.523 & \textbf{4.55} & 4.85 & Over-inclusive \\
& 11056 & 244 & 180 & 13 & 7.2\% & \textbf{0.642} & 0.607 & -- & -- & Over-inclusive \\
\bottomrule
\end{tabular}%
}
\end{table*}

\subsection{Discussion}
\noindent \textbf{Analysis of Anomaly Suppression.}
 As summarized in Tab~\ref{tab:anomaly_mask_stats}, the effect of the anomaly loss is tightly coupled to the quality of the generated pseudo-anomaly masks. On EMPIAR 10081 and 10345, where the contamination is diverse, the masks are reliable and show no mistaken overlaps with true particles, and the loss improves both F1 and average reconstruction resolution. In contrast, on EMPIAR 10093, 10532, and 11056 the loss slightly lowers F1, and our analysis reveals two distinct failure regimes.

\begin{figure*}[!t]
 \centering
 \includegraphics[width=\linewidth]{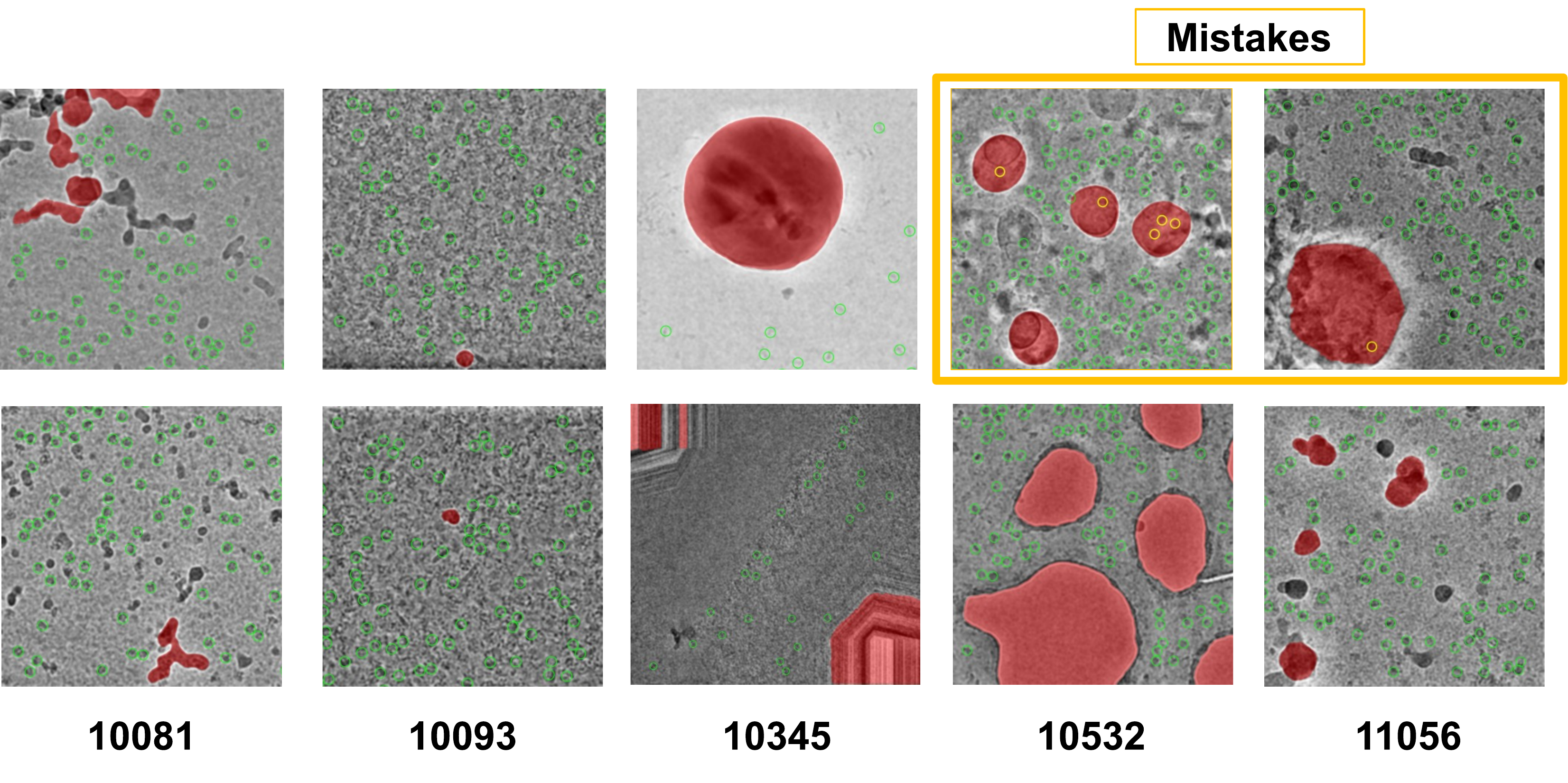}
 \caption{Visualization of generated anomaly masks across datasets. Red regions indicate generated anomaly masks. Green circles indicate Ground Truth particles. Yellow circles highlight particles that are mistakenly overlapped by anomaly masks.}
 \label{fig:anomaly_masks}
\end{figure*}

First, in datasets with relatively clean backgrounds, the anomaly detector fires on only a small fraction of training images; for EMPIAR 10093, only 3 of 236 training images receive anomaly masks. This too-sparse regime provides almost no negative supervision, rendering the anomaly loss ineffective and introducing training instability. Second, on EMPIAR 10532 and 11056 the detector is over-inclusive: 38 of 188 masked images on EMPIAR 10532 (20.2\%) and 13 of 180 masked images on 11056 (7.2\%) erroneously overlap with valid ground-truth particles. As highlighted by the yellow circles in Fig~\ref{fig:anomaly_masks}, this overlap forces the model to incorrectly penalize true particle regions as background anomalies during training, increasing false negatives.

\noindent \textbf{Limitations.}
The erroneous overlaps observed in our failure analysis highlight a limitation in our current anomaly mask generation pipeline. Currently, we rely on the Segment Anything Model (SAM)~\cite{SAM} to refine the coarse anomaly predictions generated by PaDiM~\cite{PaDiM}. However, because SAM is a general-purpose segmentation model without specific knowledge of cryo-EM physics, it can produce inaccurate mask boundaries that inadvertently encompass true particles. Replacing this generic post-processing step with a domain-specific network tailored for cryo-EM artifact definition remains a key area for future improvement.

\section{Future Work}
\label{sec:2d_to_3d}
Current particle picking methods, including ours, primarily optimize 2D metrics like F1 score. However, since the ultimate objective is 3D reconstruction, exhaustive detection of every particle is not strictly necessary. Furthermore, manual annotations in benchmarks like CryoPPP are not always exhaustive, implying that comparisons based solely on 2D scores can be ambiguous. Consequently, it is more critical to prioritize retrieving particles with diverse viewing angles while strictly excluding contaminants.
In future work, we will leverage synthetic data to quantitatively analyze the impact of particle quantity, contamination levels, and angular distribution on the final 3D map. This analysis relies on ground-truth projection angles provided by synthetic data and the controlled contamination handling as demonstrated in CryoAnomaly. By isolating these factors, we aim to establish new metrics tailored directly for reconstruction quality, addressing the limitations of relying exclusively on 2D picking performance.

\section{Conclusion}
We present CryoAnomaly, a framework that effectively leverages synthetic data for few-shot particle picking. By utilizing clean synthetic priors to detect real-world artifacts as anomalies, we turn the Sim2Real gap into a strategic advantage for robust false positive suppression. Combined with diversity-based active learning, our approach achieves the best overall few-shot accuracy on the CryoPPP benchmark, and the anomaly-guided loss is confirmed to be effective on datasets with diverse contamination, which is consistent with the observed association between reduced contaminant picks and higher-resolution 3D reconstruction.

\section*{Acknowledgment}
This work was supported in part by U.S. NSF grants DBI-2238093, DBI-2422619, IIS-2211597, and MCB-2205148.
Illustrations of the microscope and vaccine in Fig~\ref{fig:teaser} were generated using Gemini (Google)~\cite{Gemini} based on the authors' conceptual descriptions.

\clearpage
\raggedbottom
\bibliographystyle{IEEEtran}
\bibliography{refs}
\newpage

\begin{IEEEbiography}[{\includegraphics[width=1in,height=1.25in,clip,keepaspectratio]{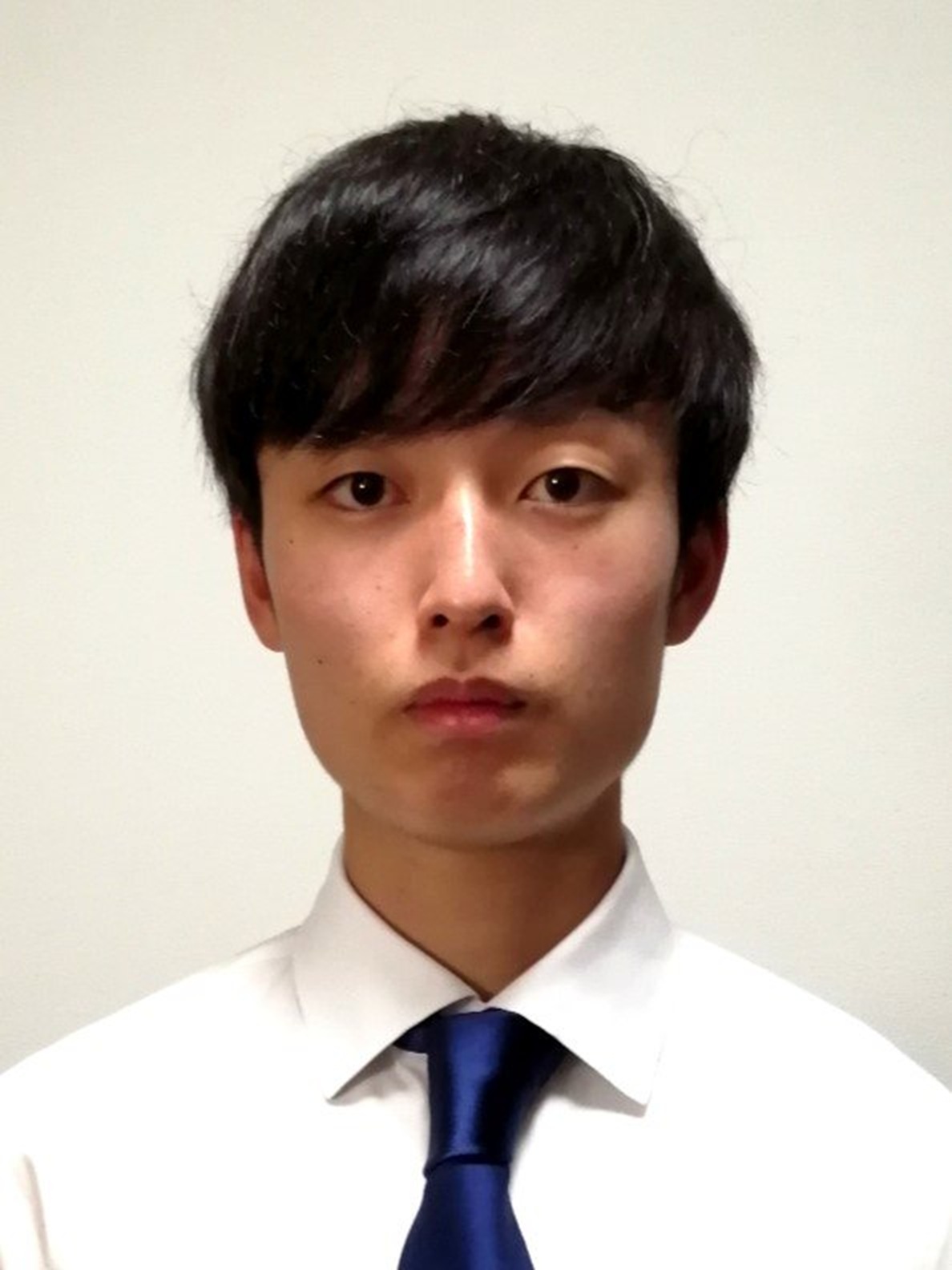}}]{Riku Itsuji} (Student Member, IEEE)
received the B.E. degree in system design engineering from Keio University, Japan,
in 2025, where he is currently pursuing the
M.Sc.Eng. degree in science and technology. His
research interests include cryo-EM image analysis and 3D reconstruction.
\end{IEEEbiography}

\begin{IEEEbiography}[{\includegraphics[width=1in,height=1.25in,clip,keepaspectratio]{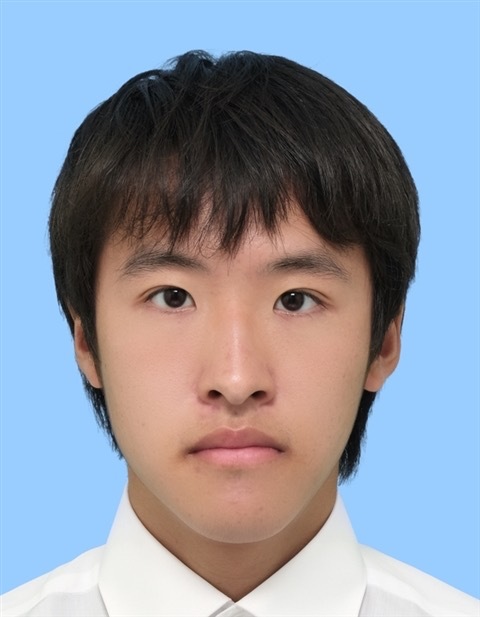}}]{Rintaro Otsubo} (Student Member, IEEE)
received the B.E. degree in information and computer science from Keio University, Japan, in 2025, where he is currently pursuing the M.Sc.Eng. degree in science and technology. His research interests include bio-medical vision analysis and transfer learning.
\end{IEEEbiography}

\begin{IEEEbiography}[{\includegraphics[width=1in,height=1.25in,clip,keepaspectratio]{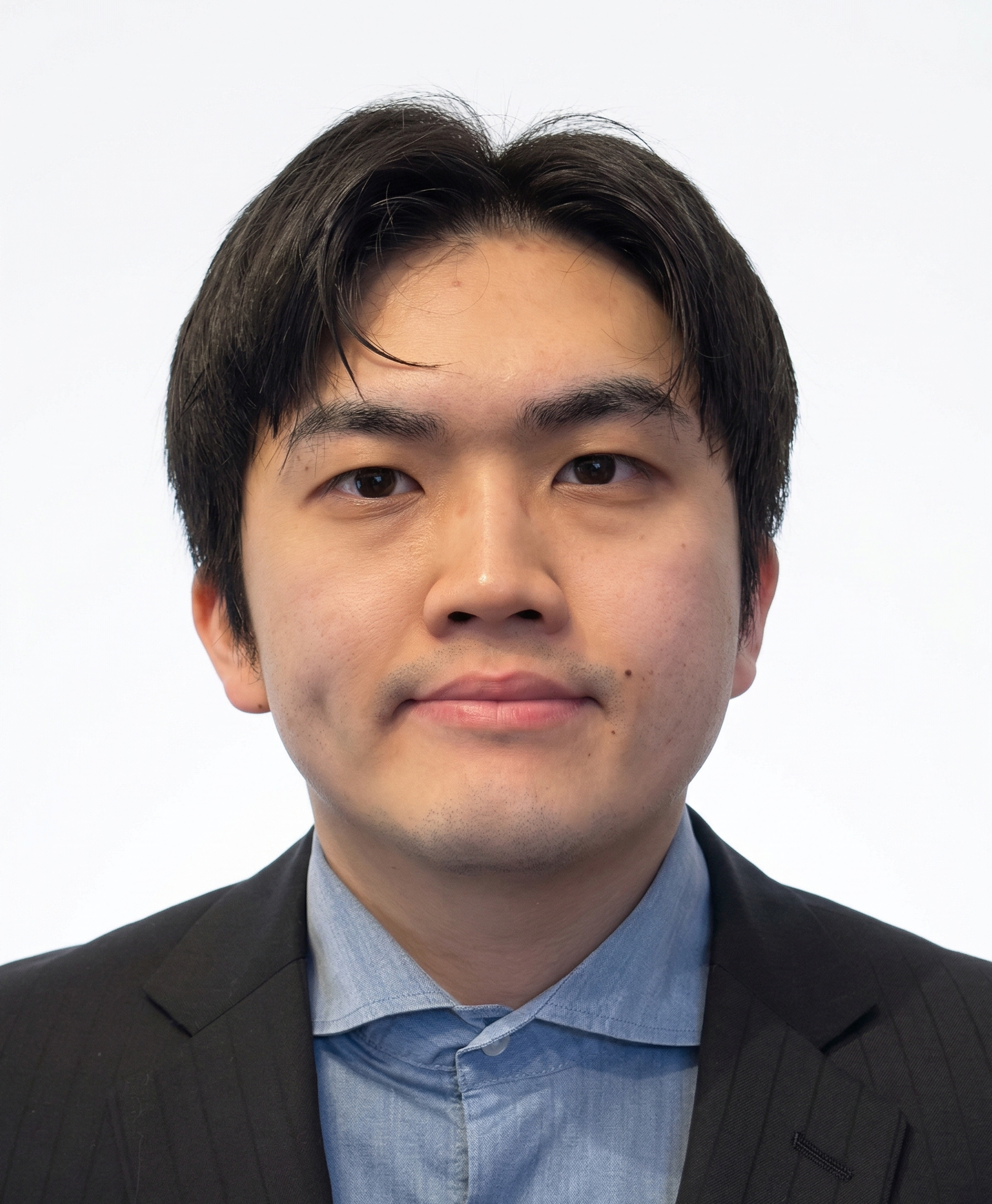}}]{Ryo Fujii}
received the B.S., M.S., and Ph.D. degrees in Engineering from Keio University, Yokohama, Japan, in 2020, 2022, and 2026, respectively. From 2025 to 2026, he was a Visiting Researcher at Carnegie Mellon University, Pittsburgh, PA, USA. His research interests include computer vision and machine learning, particularly pedestrian trajectory forecasting and video understanding, with a focus on deep learning applications in real-world environments.
\end{IEEEbiography}

\begin{IEEEbiography}[{\includegraphics[width=1in,height=1.25in,clip,keepaspectratio]{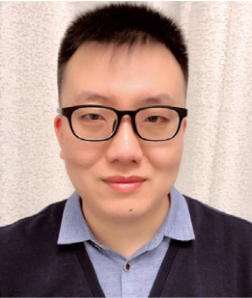}}]{Xingjian Li}
 received the B.S. degree in microelectronics from Tsinghua University, in 2008, the M.S. degree in computer science and technology from the Institute of Computing Technology, Chinese Academy of Sciences, in 2011. He received a Ph.D. degree in Computer Science at University of Macau in 2023. He is currently a Postdoctoral Associate at Carnegie Mellon University. His research interests include transfer learning, semi-supervised learning, interpretable deep learning and so on.
\end{IEEEbiography}

\begin{IEEEbiography}[{\includegraphics[width=1in,height=1.25in,clip,keepaspectratio]{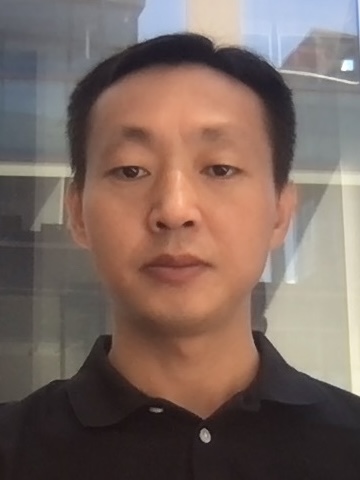}}]{Xiaolong Wu}
received his PhD in Computer Engineering from Purdue University. He is currently a Postdoctoral Associate within the School of Computer Science, Carnegie Mellon University. His research interests include computer vision, robotics, and ML system.
\end{IEEEbiography}

\begin{IEEEbiography}[{\includegraphics[width=1in,height=1.25in,clip,keepaspectratio]{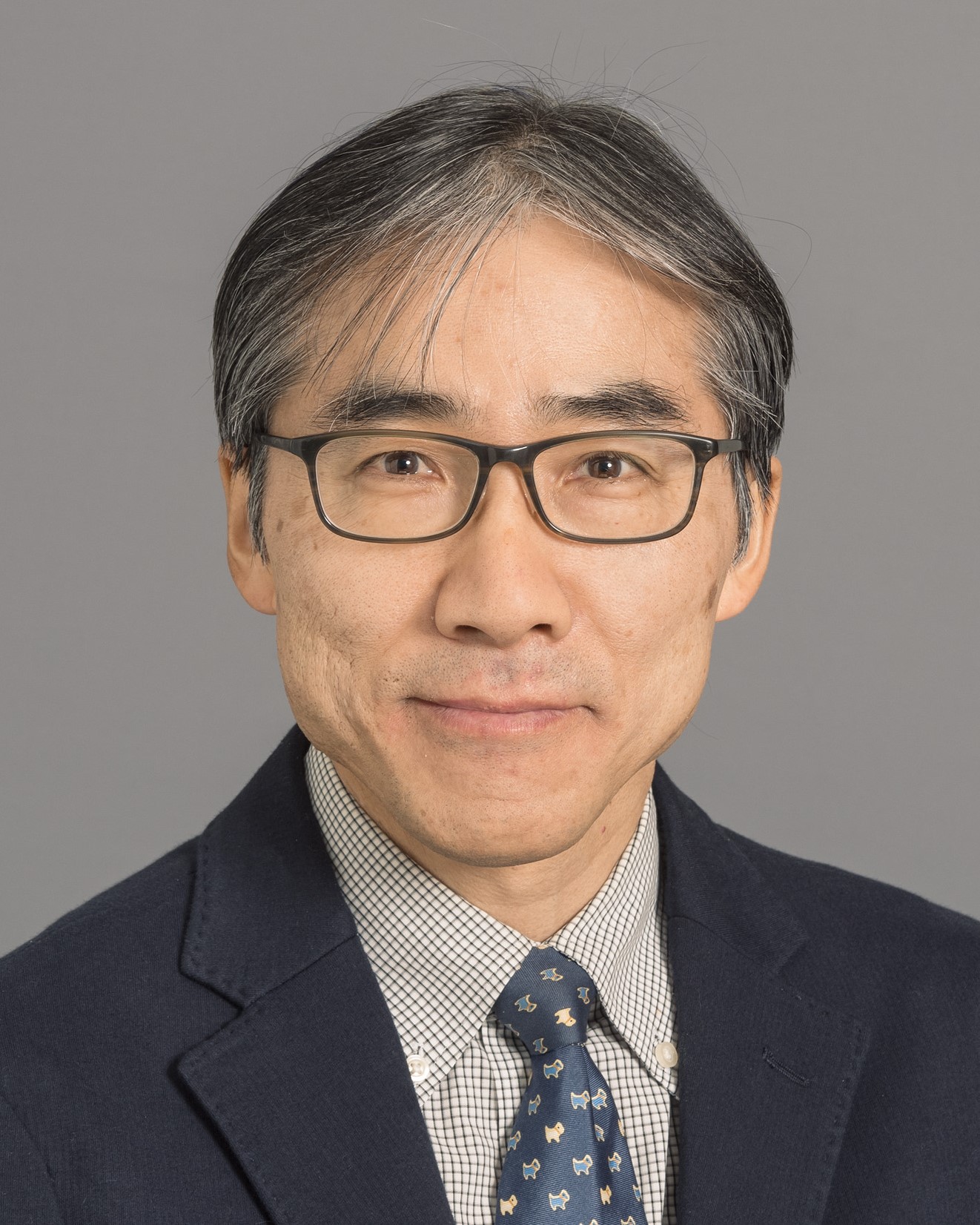}}]{Hideo Saito} (Senior Member, IEEE) received his Ph.D. degree in electrical engineering from Keio University, Japan, in 1992. Since 1992, he has been in the Faculty of Science and Technology, Keio University. From 1997 to 1999, he joined the Virtualized Reality Project at the Robotics Institute, Carnegie Mellon University, as a Visiting Researcher. Since 2006, he has been a Full Professor with the Department of Information and Computer Science, Keio University. His research interests include computer vision and pattern recognition, and their applications to augmented reality, virtual reality, and human–robotic interaction. His recent activities in academic conferences include being the Program Chair of ACCV 2014, the General Chair of ISMAR 2015, and the Program Chair of ISMAR 2016.
\end{IEEEbiography}

\begin{IEEEbiography}[{\includegraphics[width=1in,height=1.25in,clip,keepaspectratio]{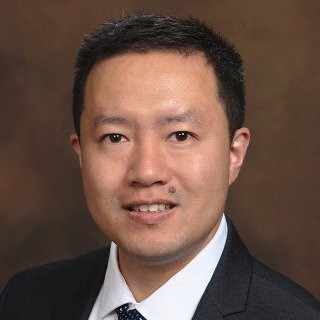}}]{Min Xu}
received the BE degree in computer science from the Beihang University, the MSc degree from School of Computing, National University of Singapore, the MA degree in applied mathematics from the University of Southern California (USC), and the PhD degree in computational biology and bioinformatics from USC. He is currently an associate professor with the Computational Biology Department within the School of Computer Science, Carnegie Mellon University.
\end{IEEEbiography}

\EOD

\end{document}